\documentclass[reprint,superscriptaddress,nobibnotes,amsmath,amssymb,prl]{revtex4-1}

\usepackage{graphicx}
\usepackage{dcolumn}
\usepackage{bm}
\usepackage{multirow}
\usepackage{mathptmx}
\usepackage{siunitx}
\usepackage{xcolor}

\newcommand{\beq}[1]{\begin{equation}\label{#1}}
\newcommand{\eep}{\;.\end{equation}}
\newcommand{\eec}{\;,\end{equation}}
\newcommand{\eeq}{\end{equation}}
\newcommand*\dd{\mathop{}\!\mathrm{d}}

\newcommand{\grad}{\nabla}
\newcommand*\chem[1]{\ensuremath{\mathrm{#1}}}

\newcommand{\lb}{\left(}
\newcommand{\rb}{\right)}

\renewcommand{\d}{\delta}
\newcommand{\ep}{\epsilon}
\newcommand{\la}{\lambda}
\newcommand{\s}{\sigma}
\newcommand{\p}{\phi}

\newcommand{\D}{\Delta}

\DeclareMathAlphabet{\mathcal}{OMS}{cmsy}{m}{n} 

\newcommand{\Ef}{\mathcal{E}}

\newcommand{\F}{\mathcal{F}}
\newcommand{\bigO}{\mathcal{O}}

\newcommand{\eo}{{\epsilon}_0}
\newcommand{\Xp}{X_{\p}}
\newcommand{\Xpi}{X_{\p}^{-1}}
\newcommand{\dc}{d_{\text{c}}}

\newcommand{\dphi}{d_{\phi}}
\newcommand{\Ec}{\mathcal{E}_{\text{c}}}

\graphicspath{{./img/}}

\begin{document}


\title{Coupling between tilts and charge carriers at polar-nonpolar perovskite interfaces}

\author{Daniel Bennett}
\email{db729@cantab.ac.uk}
\affiliation{Theory of Condensed Matter Group, Cavendish Laboratory, University of Cambridge, J.\,J.\,Thomson Avenue, Cambridge CB3 0HE, United Kingdom}
 \affiliation{Physique Th{\'e}orique des Mat{\'e}riaux, QMAT, CESAM, University of Li{\`e}ge, B-4000 Sart-Tilman, Belgium}

\author{Pablo Aguado-Puente}
\affiliation{CIC Nanogune BRTA, Tolosa Hiribidea 76, 20018 San Sebastian, Spain}

\author{Emilio Artacho}
\affiliation{Theory of Condensed Matter Group, Cavendish Laboratory, University of Cambridge, J.\,J.\,Thomson Avenue, Cambridge CB3 0HE, United Kingdom}
\affiliation{CIC Nanogune BRTA, Tolosa Hiribidea 76, 20018 San Sebastian, Spain}
\affiliation{Donostia International Physics Center, Paseo Manuel de Lardizabal 4, 20018 San Sebastian, Spain}
\affiliation{Ikerbasque, Basque Foundation for Science, 48011 Bilbao, Spain}
 
\author{Nicholas Bristowe}
\affiliation{School of Physical Sciences, University of Kent, Canterbury, Kent CT2 7NH, United Kingdom}
\affiliation{Centre for Materials Physics, Durham University, South Road, Durham DH1 3LE, United Kingdom}

\date{\today}

\begin{abstract}
  The phenomenological theory for the polar instability giving rise to 
a two-dimensional electron gas at perovskite interfaces is hereby 
extended to include the coupling to perovskite tilts.
  A Landau theory for homogeneous tilts is first explored, setting
the scene for the further, more realistic Landau-Ginzburg theory
describing varying tilt amplitudes across a thin film.
  The theory is also generalized to account for the response to
an applied electric field normal to the interface, which allows
a finer control on phase transitions.
  The conventionally described physics of a single metal-insulator 
transition is substantially enriched by the coupling, the model 
describing various scenarios with one or two transitions, 
possibly continuous or discontinuous.
  First-principles calculations permit the estimation of
the parameters defining the model, which have been calculated for
the interface between lanthanum aluminate and strontium titanate.
\end{abstract}
  
\maketitle


\section{Introduction}

  The discovery of an insulator--metal transition at the 
interface between two insulating perovskites sparked considerable interest in the 
scientific community \cite{ohtomo2002,ohtomo2004}. 
  When thin films of lanthanum aluminate 
(\chem{LaAlO_3}, LAO) were grown on substrates of strontium titanate 
(\chem{SrTiO_3}, STO), a two-dimensional electron gas (2DEG) was 
found to appear at the interface in order to screen the polar 
discontinuity there. 
  This 2DEG has been found to be associated with interesting 
phenomena such as enhanced capacitance 
\cite{2deg_enhanced_capacitance}, superconductivity \cite{2deg_superconductivity} 
and magnetism \cite{2deg_magnetism}, even at the same time 
\cite{2deg_coexistence_1,2deg_coexistence_2}. 
  It also has potential for applications in field effect 
transistors (FET) \cite{tunable_electrostatic,cen2008nanoscale,
forg2012field}, sensors \cite{2deg_sensor}, photodetection
\cite{2deg_photodetection}, thermoelectrics 
\cite{2deg_thermoelectric_1,2deg_thermoelectric_2} and solar cells 
\cite{2deg_solar_1,2deg_solar_2}.

  The physical origin and character of this insulator--metal 
transition have been debated for many years, and discussed in 
several reviews \cite{hwang2006tuning,pauli2008conducting,
huijben2009structure,mannhart2010oxide,pentcheva2010electronic,
chen2010electronic,warusawithana2013laalo,chambers2011understanding,
pentcheva2012termination,gabay20132,stemmer2014two,
bristowe_electronic_reconstruction}.
In response to a polar discontinuity at the
interface, first proposed in the original papers on LAO/STO
\cite{ohtomo2002,ohtomo2004}, two of the most popular 
theories for the appearance of the 2DEG
are indirectly supported by experimental evidence: the first 
is electronic reconstruction, where the 2DEG forms via a 
transfer of electrons from the valence band at the surface 
of the thin film to the conduction band at the interface 
in order to screen the polar discontinuity 
\cite{nakagawa2006some,popovic2008origin,pentcheva2009avoiding}. 
  The second achieves the same screening by means of redox 
defects, where oxygen vacancies or 
hydrogen adatoms, for example, form at the surface, creating 
charge carriers that can move to the interface
\cite{kalabukhov2007effect,siemons2007origin,herranz2007high,
eckstein2007watch,shibuya2007metallic,chen2011metallic,superlattice_defect,lee2012creation}. 
  In Ref.~\onlinecite{coey} it was predicted that both 
mechanisms are possible, depending on the experimental 
conditions such as the oxygen pressure and the growth 
temperature \cite{lee2012creation,temperature_experiment}.

The character of the transition with film thickness, i.e. whether the carriers appear continuously or discontinuously after a critical thickness, is still debated. It has been observed in experimental studies \cite{ohtomo2004} and suggested in theoretical studies \cite{bristowe_electronic_reconstruction,zunger_dft_paper} based on redox effects that a discontinuous transition occurs at a critical thickness $\dc$ between 3 and 4 unit cells of LAO. However, film thickness is not a good parameter with which to make any conclusions about the order of the transition, since it is fixed for each sample and is changed discretely by a number of unit cells. Thus, it is impossible to conclude whether the transition is continuous or discontinuous, and the problem is only of theoretical interest. A more realistic approach to investigate this would be to apply an electric field, using top and back gates \cite{chen2016dual}, to a sample which is close to the critical thickness. An electric field can be used to switch the 2DEG on \cite{kim2015electric} and off \cite{hosoda2013transistor} in a single sample, which is a desirable feature in practical applications. It can also enhance the properties of the 2DEG such as the superconductivity \cite{bert2012gate,eerkes2013modulation,davis2018anisotropic} magneto-transport \cite{liu2015magneto} and optical behavior \cite{lei2014visible}. An electric field could be tuned with more precision and thus would be more suitable for studying the character of the insulator--metal transition experimentally.

A phenomenological theory at the mean-field level, which treats the carriers as a homogeneous charge distribution $\s$, predicts that the onset of carriers with thickness is continuous, with a critical exponent for $\sigma$ of 1 \cite{bristowe_electronic_reconstruction}. When thinking about the redox defects proposed in Ref.~\onlinecite{bristowe_electronic_reconstruction}, an assumption of non-interacting defects suggests a drastically discontinuous transition, switching on directly to $\sigma \sim P_s$ \cite{zunger_dft_paper}, $P_s$ being the polar discontinuity, half an electron per primitive unit cell surface area for LAO/STO; when it becomes favorable for one vacancy-carrier pair to form, it is favorable for all of them to form, giving the mentioned discontinuity at $\dc$. One can go beyond the mean-field level by considering the interactions between the traps \cite{superlattice_defect}. The vacancy at the surface and carrier at the interface act like a dipole, and thus there would be dipole-dipole interactions between the vacancies. This predicts a transition which is still continuous, but with a critical exponent of $\frac{2}{3}$. 

Something that has to our knowledge not yet been considered is the influence of other phase transitions, such as the antiferrodistortive (AFD) rotations of oxygen octahedra (tilts), which both LAO and STO can exhibit. Tilts compete with the polar mode, so it is reasonable to expect that they may indirectly interact with the carriers, which appear to screen the polar discontinuity. The competition between tilts and the polar mode in LAO/STO has been considered in a previous study \cite{stengel_tilt}, although the influence of the tilts on the appearance of the carriers at the phenomenological level has not been considered.

In this paper we generalize the phenomenological model of carrier formation at polar-nonpolar perovskite interfaces to account for coupling to tilts in the thin film. We show that, upon coupling to homogeneous tilts, four new distinct types of transitions are possible, depending only on the energetics of the tilts, polar discontinuity and the coupling between them. These include continuous and discontinuous transitions of the carriers, facilitated by the tilts, and both simultaneous and distinct transitions of tilts and carriers. Using first-principles calculations, we can make predictions about the type of transition which occurs at the LAO/STO interface. We then generalize the model to allow for inhomogeneous tilts in the polar thin film, using Ginzburg-Landau theory. The inhomogeneity of the tilts is determined by the correlation length in the film and the extrapolation lengths, which describe the relative energy differences between the tilts in the interior and at the surfaces. We show that two of the transitions predicted at the homogeneous level are unaffected by the inhomogeneity of the tilts, other than the values at which the transitions occur being renormalized. For the other two transitions, their character can be changed by decreasing the correlation length, leading to two entirely new types of transitions which are not possible for homogeneous tilts.


\section{Coupling to homogeneous tilts}

\subsection{Phenomenological theory of carrier formation}

\begin{figure}[ht] 
\centering
\includegraphics[width=\columnwidth]{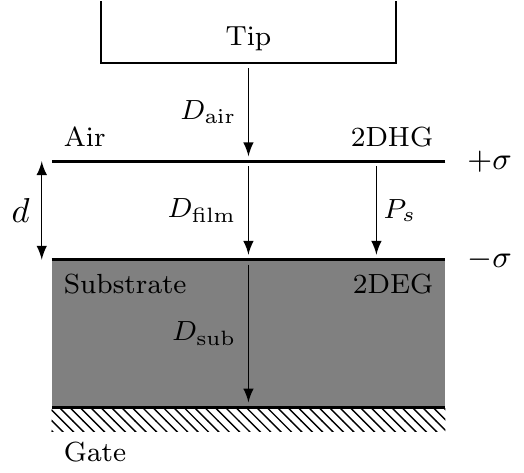}
\caption{Sketch of the electrostatics of a polar-nonpolar interface. A thin film of thickness $d$ is placed on top of a semi-infinite substrate. An electric field is applied using a back gate under the substrate and a biased tip. For simplicity, we assume that the tip is much wider than the thickness of the film, and that the applied field is homogeneous. A 2DEG forms at the interface and a corresponding 2DHG or redox defects form at the surface of the thin film as the combined result of the polar discontinuity and applied field. $D_{\text{film}}$ is the displacement field of the film and $P_s$ is the polar discontinuity. $D_{\text{air}}$ is the displacement field of the region of air between surface of the film and the tip, and $D_{\text{sub}}$ is the displacement field of the substrate.}
\label{fig:ch-3-film}
\end{figure}

First, we review the phenomenological model of carrier formation at polar-nonpolar perovskite interfaces \cite{bristowe_electronic_reconstruction}, introducing the applied electric field in a similar manner to Ref.~\cite{pablo_2deg_theory}. A sketch of the system is shown in Fig.~\ref{fig:ch-3-film}. We have a polar thin film of thickness $d$ on top of a non-polar substrate, with e.g.~air above, similar to the geometry considered for ferroelectric thin films in Ref.~\cite{bennett2020electrostatics}. Experimentally, a potential bias is usually applied between the tip of an atomic force microscope (AFM) at the surface of the film and a back gate at the bottom of the substrate \cite{caviglia2008electric,stengel2009electric,forg2012field,hosoda2013transistor}. Realistically, this would result in inhomogeneous electric fields which decay away from the tip. However, we note for example that, in the case of fields applied with an AFM tip, the lateral extent over which the field can be considered homogeneous ($\sim 100$ of nm) is much larger than the usual thickness of the polar film (few nm). This is at least in part due to the meniscus of water and other surface adsorbates that form around the AFM tip \cite{paruch2013nanoscale}. In the interest of simplicity we consider an homogeneous field, assuming the tip is much wider than the thickness of the film. With the usual LaO/TiO$_2$ termination, a 2DEG of carrier concentration $\s$ forms between the film and substrate -- and a corresponding two-dimensional hole gas (2DHG) or redox defects form at the surface -- in order to screen the polar discontinuity.

The displacement fields in the film, substrate and air region between the film and tip are $D_{\text{film}} = \ep_{\text{film}} \Ef_{\text{film}} + P_s$, ${D_{\text{sub}} = \ep_{\text{sub}}} \Ef_{\text{sub}}$, and $D_{\text{air}} = \ep_{\text{air}} \Ef_{\text{air}}$, respectively, where $P_s$ is the polar discontinuity. We consider the displacement field in the substrate as a tunable parameter with which the field (or potential difference) applied to the system is measured. The displacement field in the film could also be used \cite{stengel_tilt}. All vectors are normal to the interface, so vector notation is omitted. From Maxwell's equations, the boundary condition at interface relates the displacement fields in the film and the substrate to the free charge, $\s$:
\beq{eq:ch-3-D-boundary-condition}
\begin{gathered}
\lb D_{\text{sub}} - D_{\text{film}} \rb \cdot \hat{n} = -\s \\
\ep_{\text{sub}} \Ef_{\text{sub}} - \ep_{\text{film}} \Ef_{\text{film}} - P_s = -\s \\
\implies \Ef_{\text{film}} = \frac{\ep_{\text{sub}}}{\ep_{\text{film}}}\Ef_{\text{sub}} - \frac{1}{\ep_{\text{film}}}(P_s - \s)
\end{gathered}
\eec
where $\hat{n}$ points outwards from the film. Assuming that the boundaries of the film have equal and opposite carrier concentration, the boundary condition at the surface between the film and the air region is also satisfied by enforcing finite-$D$ boundary conditions: $D_{\text{sub}} = D_{\text{air}}$ (not the electric fields).

\begin{figure}[ht!] 
\centering
\includegraphics[width=\columnwidth]{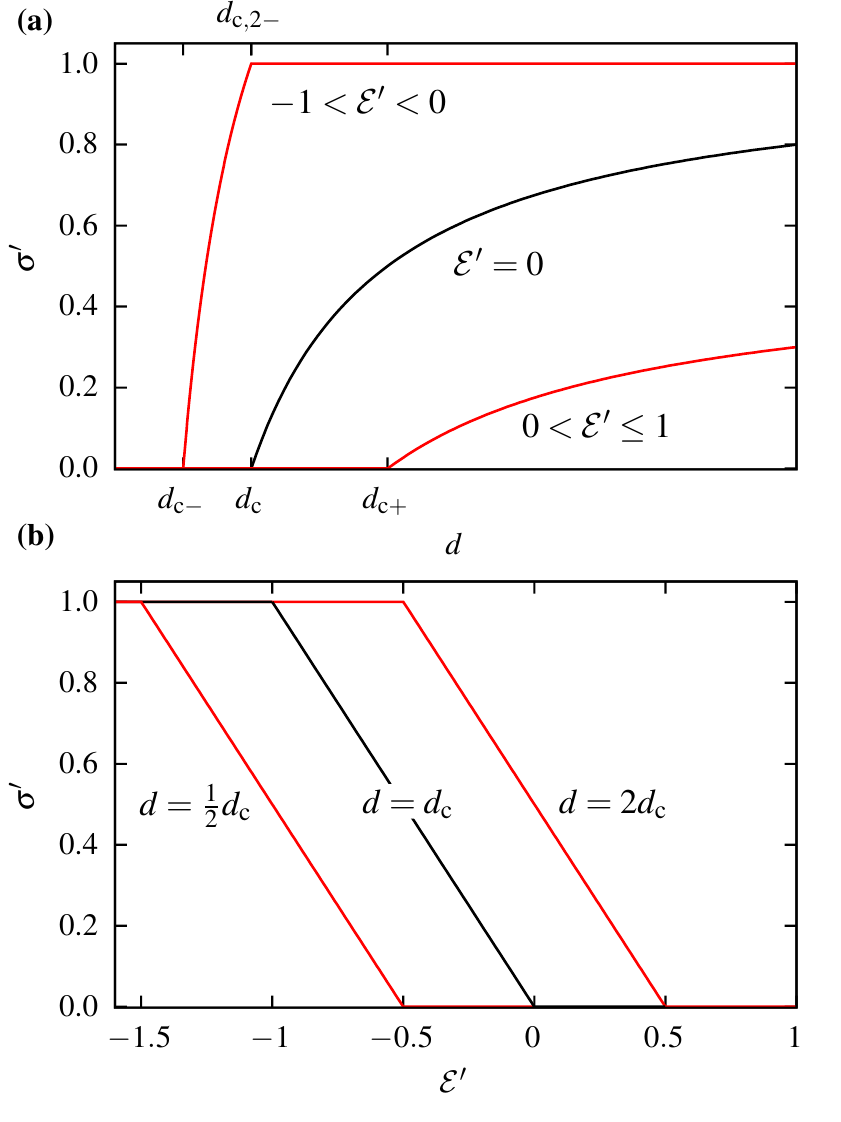}
\caption{\textbf{(a)}: Carrier concentration as a function of thickness for zero (black), positive and negative (red) values of reduced field $\Ef'$. When $\Ef'$ is positive, $\dc$ increases to $d_{\text{c}+}$. When $\Ef'$ is negative, $\dc$ is reduced to $d_{\text{c}-}$, and $\s'$ saturates to 1 at a second critical thickness. \textbf{(b)}: Reduced carrier concentration with applied field at different thicknesses. The critical thickness $d=\dc$ is shown in black, and the sub- and super-critical thicknesses, $d=\frac{1}{2}\dc$ and $d=2\dc$, respectively, are shown in red.}
\label{fig:ch-3-sigma-d-E}
\end{figure}

The electrostatic energy of the film is \cite{jackson1999classical}
\beq{eq:ch-3-F-elec-1}
\F_{\text{elec}} = \int_{0}^{D_{\text{film}}} \Ef_{\text{film}} \cdot \d D_{\text{film}} =  \frac{1}{2}\ep_{\text{film}}\Ef_{\text{film}}^2
\eep
Using Eq.~\eqref{eq:ch-3-D-boundary-condition} we can write the electrostatic energy in terms of $\Ef_{\text{sub}}$, $P_s$ and $\s$:
\beq{eq:ch-3-F-elec-1}
\F_{\text{elec}} = \frac{1}{2}\frac{\ep_{\text{sub}}^2}{\ep_{\text{film}}}\Ef_{\text{sub}}^2 -\frac{\ep_{\text{sub}}}{\ep_{\text{film}}}\Ef_{\text{sub}}\cdot(P_s-\s) + \frac{1}{2\ep_{\text{film}}}(P_s-\s)^2
\eeq
The first term corresponds to the energy of the applied field in the substrate, which does not affect the carrier concentration and can be neglected (the energy of the substrate is also neglected for the same reason). The second term represents the coupling between the applied field and the polar mode and the third term is the electrostatic energy of the polar discontinuity. Writing Eq.~\eqref{eq:ch-3-F-elec-1} in terms of reduced variables $\s'=\frac{\s}{P_s}$ and $\Ef'=\frac{\Ef_{\text{sub}}}{(P_s/\ep_{\text{sub})}}$, which is the displacement field in the substrate, in units of the polar discontinuity, we get
\beq{eq:ch-3-F-elec-2}
\F_{\text{elec}}(\s',\Ef') = \frac{P_s^2}{2\ep}(1-\s')^2 -\frac{P_s^2}{\ep}\Ef'\cdot(1-\s')
\eec
where we drop the subscript $\ep_{\text{film}} \to \ep$, since the permittivity of the substrate has been absorbed into the scale for the applied field and no longer appears in the free energy. For $\Ef' = 0$, the energy is minimized when $\s' = 1$, i.e. the polar discontinuity is fully screened by the carriers. For LAO/STO, this corresponds to a carrier concentration of exactly half an electron per unit cell surface area. However, we have neglected the formation energy of the carriers which appear the interface. Including the cost of generating an electron-hole pair across a gap $\D$, the free energy is
\beq{eq:ch-3-F-sigma-1}
\F_{\s}(\s',\Ef') = \frac{P_s^2}{\ep}\left[\frac{\dc}{d}\s' + \frac{1}{2}(1-\s')^2 -\Ef'\cdot(1-\s') \right]
\eec
where $d_c = \frac{\eo \D}{P_s}$ is the critical thickness, above which it is favorable for carriers to appear. Minimizing with respect to $\s'$ gives:
\beq{eq:ch-3-sigma-dE-solution}
\s'(d,\Ef') = 1 - \lb \frac{\dc}{d}+\Ef'\rb
\eec
which implies that the carrier concentration increases with the thickness of the film, and can be tuned, and even switched on/off with an electric field. When an electric field is applied, the critical thickness for carriers to appear is
\beq{eq:ch-3-d-crit-E}
\dc(\Ef') = \frac{\dc}{1-\Ef'}
\eec
which can be reduced or increased, depending on the sign of $\Ef'$; the sign convention used for electric field is so that it is aligned with the polar discontinuity when positive. In Fig.~\ref{fig:ch-3-sigma-d-E} (a) we show the carrier concentration as a function of thickness at zero field, and at positive and negative values of $\Ef'$. When $\Ef'=0$, the appearance of carriers is as described in Ref.~\onlinecite{bristowe_electronic_reconstruction}: $\s'$ switches on at $\dc$ and approaches 1 asymptotically from below. When a negative field is applied, the polar discontinuity enhanced, and the carrier transition occurs at a reduced thickness. In this case there is a second critical thickness where $\s'$ reaches 1 and saturates:
\beq{}
d_{\text{c},2}(\Ef') = -\frac{\dc}{\Ef'}
\eep

When $\Ef' = -1$, the critical thicknesses are $\dc(-1) = \frac{\dc}{2}$ and $d_{\text{c},2}(-1) = \dc$, i.e. the carriers appear at half the original critical thickness and saturate at the original critical thickness. When a positive field is applied, the polar discontinuity is partially screened, and the critical thickness increases.

The other possibility is to fix the thickness and induce the carriers with an applied field. From Eq.~\eqref{eq:ch-3-d-crit-E}, at a fixed thickness there is a critical field value for which the carriers appear:
\beq{}
\Ec'(d) = 1-\frac{\dc}{d}
\eep
If $d < \dc$, the film is sub-critical and $\Ec'(d)<0$. If $d \geq \dc$, the carriers will already be present at zero field. Applying a positive field partially screens the polar discontinuity, reducing the need for carriers. There is a second critical field value at which the carrier concentration saturates:
\beq{}
\Ef_{\text{c},2}'(d) = -\frac{\dc}{d}
\eec
which is always negative for finite $d$. We have the relation between the two critical field values:
\beq{}
\Ec'(d) - \Ef_{\text{c},2}'(d) = 1
\eep
The different scenarios for inducing or suppressing carriers with an applied field are summarized in Fig.~\ref{fig:ch-3-sigma-d-E} (b).

\subsection{Coupling to tilts}

In this section we consider the effect of coupling to homogeneous tilts on the formation of carriers. The simplest way to consider this is to add the independent free energies of the tilts and carriers plus a coupling term: $\F = \F_{\s} + \F_{\p} + \F_{\s\p}$.

Bulk LAO undergoes a transition from cubic to rhombohedral at $T_C \sim 541 ^{\circ}\si{C}$ \cite{lehnert2000powder}. The tilt pattern observed is, in Glazer notation \cite{glazer}, $a^-a^-a^-$. For a thin film of LAO, the tilt pattern changes to $a^-a^-c^0$ due to the tensile biaxial strain imposed by clamping to the STO substrate \cite{tilt_strain,fister2014octahedral}. If a compressive biaxial strain were applied by using a substrate with a smaller lattice constant than LAO, the observed tilt pattern changes to $a^0a^0c^{-}$ \cite{tilt_strain}. For simplicity we assume that the thin film undergoes a simple displacive transition from untilted to tilted below some temperature $T_C$. Thus, the free energy of tilts can be described by a double well:
\beq{eq:ch-3-F-tilt-1}
\F_{\p}(\p',T) = \frac{1}{2}\Xpi\left[ \frac{1}{4}\p'^4-\frac{1}{2}\lb 1-\frac{T}{T_C}\rb\p'^2\right]
\eec
where $\p' = \frac{\p}{\p_0}$ is the tilt angle in units of $\p_0$, the bulk equilibrium tilt angle, and $\Xpi$ is the curvature about the minima of the double well, in units of energy per unit volume, playing the role of an inverse susceptibility.

The simplest coupling we can introduce is a biquadratic coupling between the tilts and the polar mode \cite{benedek2013there,carpenter_landau_coupled,stengel_tilt}:
\beq{eq:ch-3-F-coupled-1}
\F_{\s\p} = \frac{1}{2}A \p'^2 \lb\frac{\Ef_{\text{film}}(\s')}{P_s/\ep}\rb^2
\eec
where $A$ is the biquadratic coupling coefficient, in units of energy per unit volume. Combining Eqs.~\eqref{eq:ch-3-F-sigma-1}, \eqref{eq:ch-3-F-tilt-1} and \eqref{eq:ch-3-F-coupled-1}, the total free energy is
\beq{eq:ch-3-F-tot-reduced-1}
\begin{gathered}
\F_{\text{tot}} = \frac{P_s^2}{\ep}\left[\frac{\dc}{d}\s' + \frac{1}{2}\lb 1+ \frac{A}{P_s^2/\ep} \p'^2\rb \lb 1-\s'-\Ef' \rb^2\right]\\
+ \frac{1}{2}\Xpi\left[ \frac{1}{4}\p'^4-\frac{1}{2}\lb 1-\frac{T}{T_C}\rb\p'^2\right]
\end{gathered}
\eep
Note that there are three independent energy scales $\Xpi$, $\frac{P_s^2}{\ep}$ and $A$, corresponding to the tilts, the polar discontinuity, and the biquadratic coupling, respectively.

Setting $T=0$ and minimizing Eq.~\eqref{eq:ch-3-F-tot-reduced-1} with respect to $\s'$ and $\p'$, we get
\beq{eq:ch-3-solutions-reduced}
\begin{split}
\p' & = \sqrt{1 - 2AX_{\p}(1-\s'-\Ef')^2}\\
\s' &= 1-\Ef'-\frac{\dc/d}{1+\frac{A}{(P_s^2/\ep)}\p'^2}
\end{split}
\eep
Note that when we set $A=0$, we recover the solutions of the uncoupled order parameters: Eq.~\eqref{eq:ch-3-sigma-dE-solution} for $\s'$ and $\p'=1$. Eq.~\eqref{eq:ch-3-solutions-reduced} is a pair of coupled equations which does not have analytic solutions. However, we can use physical constraints to understand the behavior of $\s'$ and $\p'$ upon coupling. Firstly, the square of the tilts cannot be negative:
\beq{eq:ch-3-sigma-condition}
\begin{gathered}
\p'^2 = 1 - 2AX_{\p}(1-\s'-\Ef')^2 \geq 0\\
\implies 1-\Ef'-\frac{1}{\sqrt{2AX_{\p}}} \leq \s' \leq 1-\Ef'+\frac{1}{\sqrt{2AX_{\p}}}
\end{gathered}
\eep
Note that Eq.~\eqref{eq:ch-3-sigma-condition} provides bounds on $\Ef'$, since the upper bound must be greater than or equal to 0, and the lower bound must be less than or equal to 1:
\beq{}
-\frac{1}{\sqrt{2AX_{\p}}} \leq \Ef' \leq 1+\frac{1}{\sqrt{2AX_{\p}}}
\eec
although the upper bound is redundant because ${\s'(\Ef' > 1) < 0}$, which is unphysical. The lower bound in Eq.~\eqref{eq:ch-3-sigma-condition} can also lead to situations where $\s' < 0$. This leads to a change in behaviour when the following condition relating the coefficients $\Xpi$ and $A$ is held:
\beq{eq:ch-3-straight-line}
A = \frac{1}{2(1-\Ef')^2}\Xpi
\eep
Above this line, for $\s'=0$ we would have $\p^2 < 0$, so the tilts must be zero in the absence of carriers. Thus, the carriers will appear at $\dc$ as in the uncoupled model, and the tilts will appear at some thickness, which we call $\dphi$, which is greater than $\dc$. Below this line, there are finite tilts in the absence of carriers. In this case, carriers appear earlier at $\dphi < \dc$, facilitated by a change in the tilts. 

To summarize, above the straight line in Eq.~\eqref{eq:ch-3-straight-line}, there are two possibilities: an uncoupled appearance of carriers at $\dc$ and a kink in the carrier concentration facilitated by the appearance of tilts at $\dphi > \dc$. Below the line, both order parameters appear simultaneously at $\dphi < \dc$, i.e. there is one critical thickness which is reduced by tilting.

\begin{figure}[ht!] 
\centering
\includegraphics[width=\columnwidth]{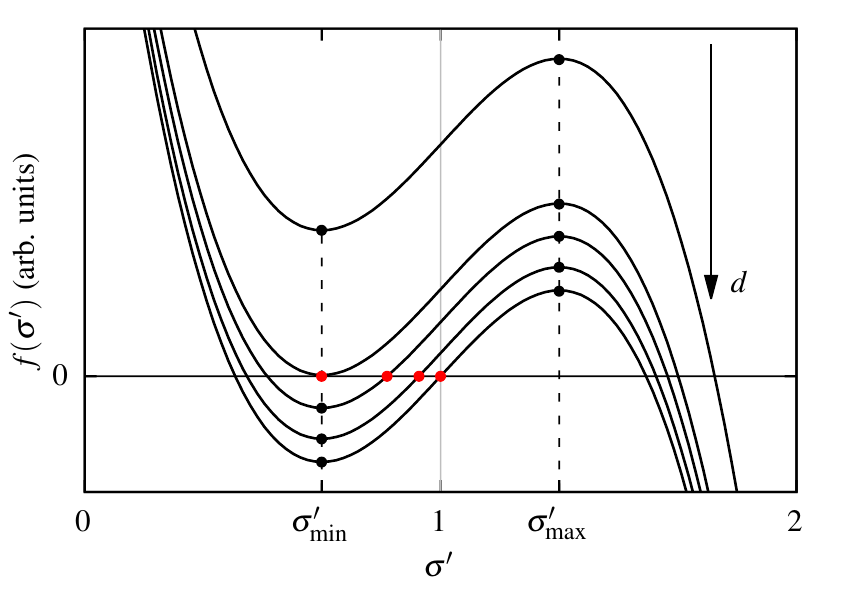}
\caption{Sketch of Eq.~\eqref{eq:ch-3-cubic} for different values of thickness. The black dots mark the maxima/minima of the curve, i.e.~Eq.~\eqref{eq:ch-3-sigma-min-max}. A non-zero carrier concentration is obtained when the polynomial has a root between 0 and 1, indicated by the red dots. For the highest curve, there is no root between 0 and 1, so the $\s=0$. When $\s'_{\text{min}}$ crosses the zero axis, we will have $\s > 0$. Since $\s'_{\text{min}} > 0$, $\s$ jumps from zero to a positive value, and the transition is discontinuous. As the curve is shifted further downwards, the root moves to the right until it reaches 1, where it saturates.}
\label{fig:ch-3-cubic}
\end{figure}

While the uncoupled carrier transition at $\dc$ is always continuous, we can investigate whether the transition at $\dphi$ is continuous or discontinuous. If we insert the solution for $\p'$ into $\s'$ in Eq.~\eqref{eq:ch-3-solutions-reduced}, we obtain a cubic equation,
\beq{eq:ch-3-cubic}
\begin{split}
f(\s')=2AX_{\p}(1-\s'-\Ef')^3 &- \lb 1+\frac{P_s^2/\ep}{A} \rb(1-\s'-\Ef')\\
&+\lb \frac{P_s^2/\ep}{A}\rb\frac{\dc}{d} = 0
\end{split}
\eec
the roots of which determine the value of $\s'$. We cannot obtain the roots analytically, but we can plot $f(\s')$ at different thicknesses and gain some insight about the transition, see Fig.~\ref{fig:ch-3-cubic}.

Since $d$ only appears in the constant term, the curve is shifted up or down by changing thicknes, and the positions of the extrema are unaffected. The maximum and minimum are uniquely determined by the energy densities $\Xpi$, $A$ $\frac{P_s^2}{\ep}$, and the field strength $\Ef'$:
\beq{eq:ch-3-sigma-min-max}
\s'_{\text{min/max}} = 1 -\Ef' \pm \sqrt{\frac{1}{6AX_{\p}}\lb 1+\frac{P_s^2/\ep}{A} \rb}
\eep
Again, Eq.~\eqref{eq:ch-3-sigma-min-max} leads to bounds on the applied field:
\beq{}
- \sqrt{\frac{1}{6AX_{\p}}\lb 1+\frac{P_s^2/\ep}{A} \rb} \leq \Ef' \leq 1+ \sqrt{\frac{1}{6AX_{\p}}\lb 1+\frac{P_s^2/\ep}{A} \rb} 
\eec
and again the upper bound is redundant. Eq.~\eqref{eq:ch-3-sigma-min-max} provides a clear picture of how the carrier transition occurs: the curve in Eq.~\eqref{eq:ch-3-cubic} is shifted by changing $d$, and carriers appear when there is a root between 0 and 1. If $\s'_{\text{min}} < 0$, the carriers appear continuously and if $0 < \s'_{\text{min}} < 1$ they appear discontinuously. Thus, $\s'_{\text{min}} = 0$ defines a boundary between first and second order transitions:
\beq{}
6(1-\Ef')^2A^2-\Xpi\lb A + \frac{P_s^2}{\ep}\rb = 0
\eec
which is quadratic in $A$ and has one positive solution:
\beq{eq:ch-3-curve}
A = \frac{1}{12(1-\Ef')^2}\Xpi \lb 1+\sqrt{1+24(1-\Ef')^2\frac{(P_s^2/\ep)}{\Xpi}}\rb
\eep
Above this curve, the transition at $\dphi$ is discontinuous and below it is continuous. Combining Eqs.~\eqref{eq:ch-3-straight-line} and \eqref{eq:ch-3-curve} results in four different possible sequences of transitions, depending only on the energy densities $\Xpi$, $\frac{P_s^2}{\ep}$, $A$ and the field $\Ef'$. 

\subsection{Transitions with thickness: zero field}

\begin{figure}[ht!] 
\centering
\includegraphics[width=\columnwidth]{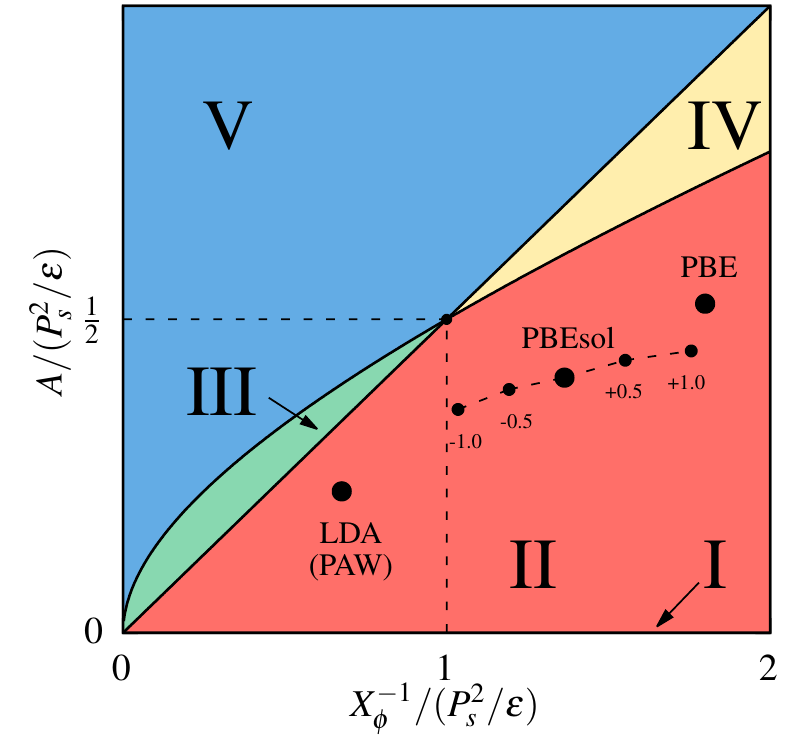}
\caption[phase2]{Diagram summarizing the coupled transitions of carriers and tilts. Region I, the line $A = 0$, describes the uncoupled order parameters. The tilts and carriers appear simulataneously in regions II (continuous) and IV (discontinuous). The tilts appear after the carriers in regions III (continuous) and V (discontinuous). First-principles calculations, summarized in Table~\ref{table:ch-3-dft}, were used to place LAO/STO on the diagram. The smaller dots show the effect of applying a small amount of biaxial strain.}
\label{fig:ch-3-phase-diagram}
\end{figure}

\begin{figure*}[ht!] 
\centering
\includegraphics[width=0.9\linewidth]{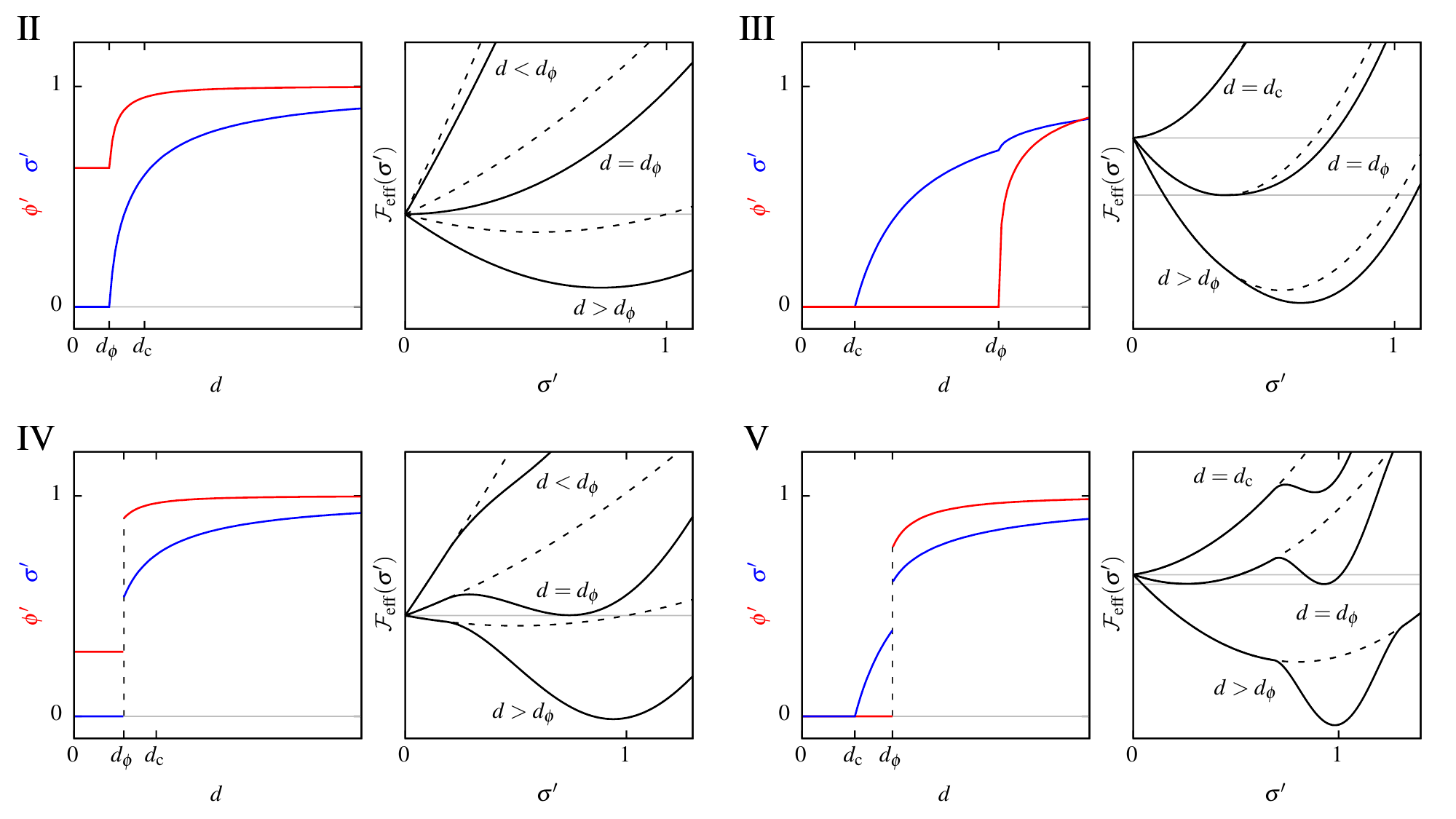}
\caption{Summary of the carrier transitions with film thickness for the four coupled regions in Fig.~\ref{fig:ch-3-phase-diagram}. The left panel for each region is an order parameter plot of $\p'$ and $\s'$ with thickness. The critical thickness from the uncoupled theory is $\dc$, and the additional critical thickness at which the tilts switch on/jump is $\dphi$. The right panel shows an effective theory, $\F_{\text{eff}}(\s')$, obtained by using Eq.~\eqref{eq:ch-3-solutions-reduced} to write Eq.~\eqref{eq:ch-3-F-tot-reduced-1} solely in terms of $\s'$. Several energy curves are shown for different thicknesses in each panel, and the dashed line shows a continuation of the uncoupled theory. The value of $\s'$ at each value of $d$ is given by the horizontal position of the minimum on each curve.}
\label{fig:ch-3-regions-thickness}
\end{figure*}

At zero field, Eqs.~\eqref{eq:ch-3-straight-line} and \eqref{eq:ch-3-curve} become

\beq{eq:ch-3-curve-zero-field}
\begin{gathered}
A = \frac{1}{2}\Xpi \\
A = \frac{1}{12}\Xpi \lb 1+\sqrt{1+24\frac{(P_s^2/\ep)}{\Xpi}}\rb
\end{gathered}
\eec

and the appearance of carriers and tilts depends only on the energy densities $\Xpi$, $\frac{P_s^2}{\ep}$, $A$. These transitions are summarized in Fig.~\ref{fig:ch-3-phase-diagram}. The two lines in Eq.~\eqref{eq:ch-3-curve-zero-field} intersect at a tetra-critical point:
\beq{eq:ch-3-tetra-zero-field}
\begin{split}
A^* &= \frac{P_s^2}{2\ep}\\
{\Xpi}^* &= \frac{P_s^2}{\ep}\\
\end{split}
\eep
In Fig.~\ref{fig:ch-3-regions-thickness}  we plot the order parameters as a function of thickness in each region of Fig.~\ref{fig:ch-3-phase-diagram}, as well as contours of the total free energy as a function of $\s'$ for several thicknesses.

\subsection{Transitions with thickness: nonzero field}

At a finite applied field, the different possible transitions with thickness are determined by the lines 

\beq{eq:ch-3-curve-nonzero-field}
\begin{gathered}
A = \frac{1}{2(1-\Ef')^2}\Xpi \\
A = \frac{1}{12(1-\Ef')^2}\Xpi \lb 1+\sqrt{1+24(1-\Ef')^2\frac{(P_s^2/\ep)}{\Xpi}}\rb
\end{gathered}
\eec

with a tetra-critical point 

\beq{eq:ch-3-tetra-nonzero-field}
\begin{gathered}
A^* = \frac{1}{2}\frac{P_s^2}{\ep}\\
{\Xpi}^* = (1-\Ef')^2\frac{P_s^2}{\ep}
\end{gathered}
\eep

\begin{figure}[ht!] 
\centering
\includegraphics[width=0.9\columnwidth]{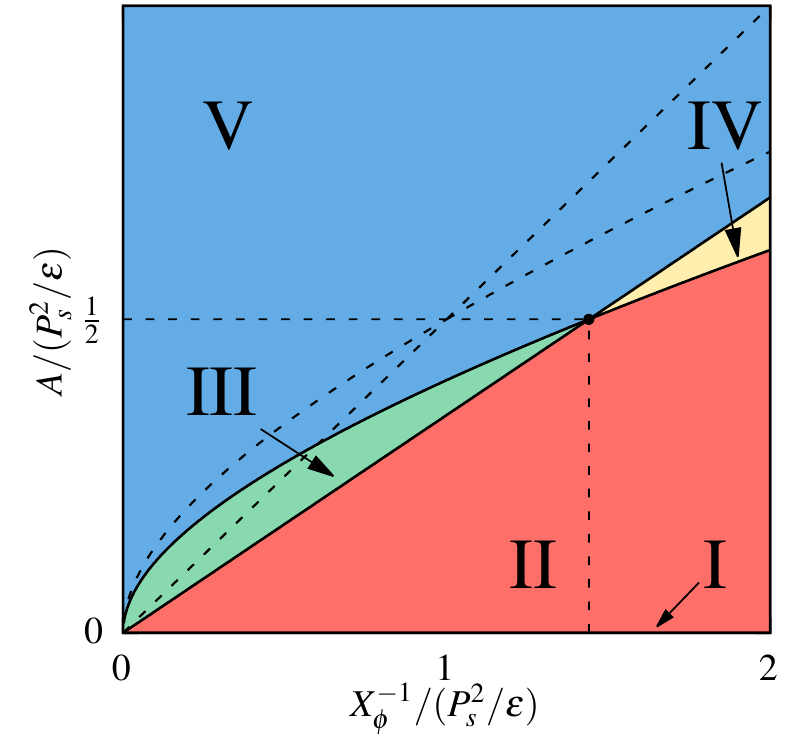}
\caption{Phase transition lines at $\Ef'=-0.2$. A positive field makes them move up and a negative field makes them move down. So it might be possible to tune the transitions with thickness by applying a fixed field. The dashed lines show the boundaries between regions at zero field.}
\label{fig:ch-3-phase-diagram-d-E}
\end{figure}

Interestingly, Eq.~\eqref{eq:ch-3-tetra-nonzero-field} reduces to Eq.~\eqref{eq:ch-3-tetra-zero-field} when $\Ef'=0,+2$, although $\Ef'=+2$ is not physical. A positive field moves the tetra-critical point to the left, making region II the largest region as $\Ef'\to 1$. A negative field moves the tetra-critical point to the right. Thus, using a negative or positive field, the boundary lines can be moved, deforming the regions, and it may be possible to tune the order of the carrier and tilt transitions with thickness at a polar-nonpolar perovskite interface. The phase transition diagram is sketched in Fig.~\ref{fig:ch-3-phase-diagram-d-E} for $\Ef'=-0.2$. We can see that the tetra-critical point moves to the right, changing the positions of the regions.

\subsection{Transitions with temperature}

In this section we investigate the possibility of inducing a carrier transition with temperature. When the order parameters are uncoupled, at a nonzero temperature we have
\beq{}
\begin{split}
\s'(d) &= 1-\frac{\dc}{d}\\
\p'(T) & = \sqrt{1-\frac{T}{T_C}}
\end{split}
\eec
where we set $\Ef' = 0$ for simplicity. For nonzero $A$, minimizing Eq.~\eqref{eq:ch-3-F-tot-reduced-1} gives
\beq{eq:ch-3-solutions-reduced_temp}
\begin{split}
\p' & = \sqrt{\lb 1-\frac{T}{T_C}\rb - 2AX_{\p}(1-\s')^2}\\
\s' &= 1-\frac{1}{1+\frac{A}{(P_s^2/\ep)}\p'^2}\frac{\dc}{d}
\end{split}
\eep
If we require that $\p^2 \geq 0$ as before, we get
\beq{}
1-\frac{1}{\sqrt{2AX_{\p}(T)}} \leq \s' \leq 1+\frac{1}{\sqrt{2AX_{\p}(T)}}
\eec
where $\Xp(T) = \lb 1-\frac{T}{T_C}\rb^{-1} \Xp$. As in Eq.~\eqref{eq:ch-3-sigma-condition}, the inequality on the left hand side can lead to situations where $\p^2 < 0$. This implies that there is an additional critical temperature, $T_C'$:
\beq{eq:ch-3-TC-reduced}
T_C' \equiv T_C \lb 1 - 2AX_{\p}\rb
\eep
This is the reduced critical temperature of the thin film, a phenomenon which is typically seen in phase transitions when going from bulk to thin films \cite{lubensky1975critical,kretschmer1979surface,
de2018superconductivity,chandra2007landau}. Interestingly, we have $T_C' = 0$ when
\beq{eq:ch-3-straight-line-temp}
A > \frac{1}{2}\Xpi
\eec
i.e.~there is no tilt transition at all. This is the same straight line obtained for transitions with thickness and applied field at zero temperature. However, it has a different meaning for transitions with temperature: below the line there is a tilt transition at a reduced critical temperature $T_C'$, and above the line there is no tilt transition. There is a second transition temperature, $T_C''$, below which the carriers appear. Inserting $\p'$ into $\s'$ in Eq.~\eqref{eq:ch-3-solutions-reduced_temp} and letting $\s'(T_C'') = 0$, we get
\beq{eq:ch-3-TC-2}
T_C'' \equiv T_C' - T_C\lb \frac{P_s^2/\ep}{A}\rb\lb \frac{\dc}{d}-1\rb
\eeq
We can see that $T_C'' < T_C'$ for sub-critical films, and $T_C'' = T_C'$ when $d=d_c$. Thus, two transitions are possible by decreasing temperature in sub-critical films and only one transition is possible otherwise. Additionally, it is possible for the transition at $T_C''$ to be first or second order. Inserting $\p'$ into $\s'$ as before, we obtain another cubic equation in $\s'$:
\beq{}
\begin{split}
f_T(\s')=\underbrace{2AX_{\p}}_{\Lambda}(1-\s')^3 &-\underbrace{\lb 1-\frac{T}{T_C}+\frac{P_s^2/\ep}{A}\rb}_{\Gamma(T)}(1-\s')\\
&+\lb\frac{P_s^2/\ep}{A}\rb\frac{\dc}{d} = 0
\end{split}
\eec
with extrema
\beq{}
\s'_{\text{max/min}}(T) = 1\pm \sqrt{\frac{\Gamma(T)}{3\Lambda}}
\eep
\begin{figure}[t] 
\centering
\includegraphics[width=\columnwidth]{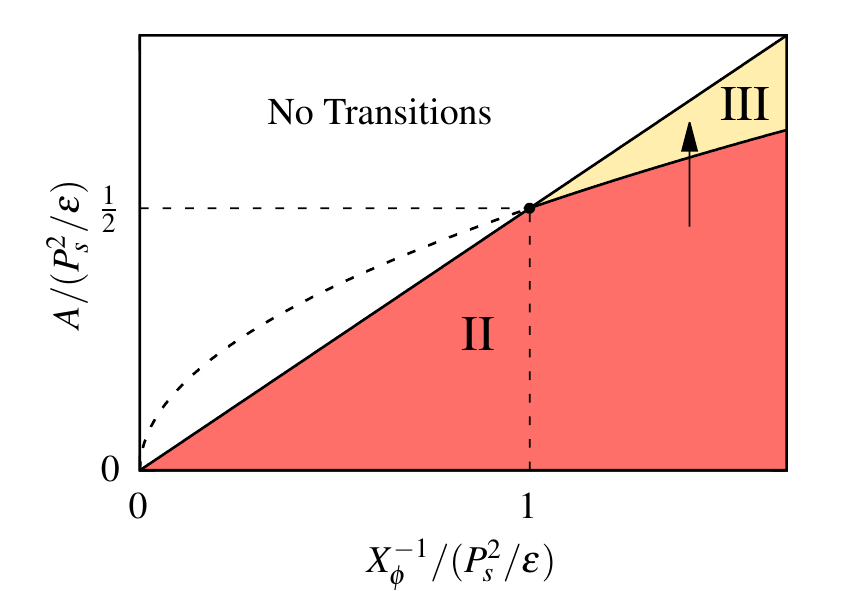}
\caption[phase2]{Phase transition diagram for transitions driven by temperature. The line $A = 0$ describes the uncoupled order parameters (region I). The straight line is the same as in Fig.~\ref{fig:ch-3-phase-diagram}. Above this line, no tilt transitions with temperature are possible: $T_C'=0$. The curved line is given by Eq.~\eqref{eq:ch-3-curved-line-temp} and determines whether the carrier transition is continuous (below) or discontinuous (above). The number of transitions, i.e.~whether or not $T_C'$ and $T_C''$ are different is determined by the ratio $\frac{d}{\dc}$ only: if $\frac{d}{\dc}< 1$ there are two transitions and if $\frac{d}{\dc} \geq 1$ there is only one transition.}
\label{fig:ch-3-phase-diagram-temp}
\end{figure}

\begin{figure*}[t!] 
\centering
\includegraphics[width=0.9\linewidth]{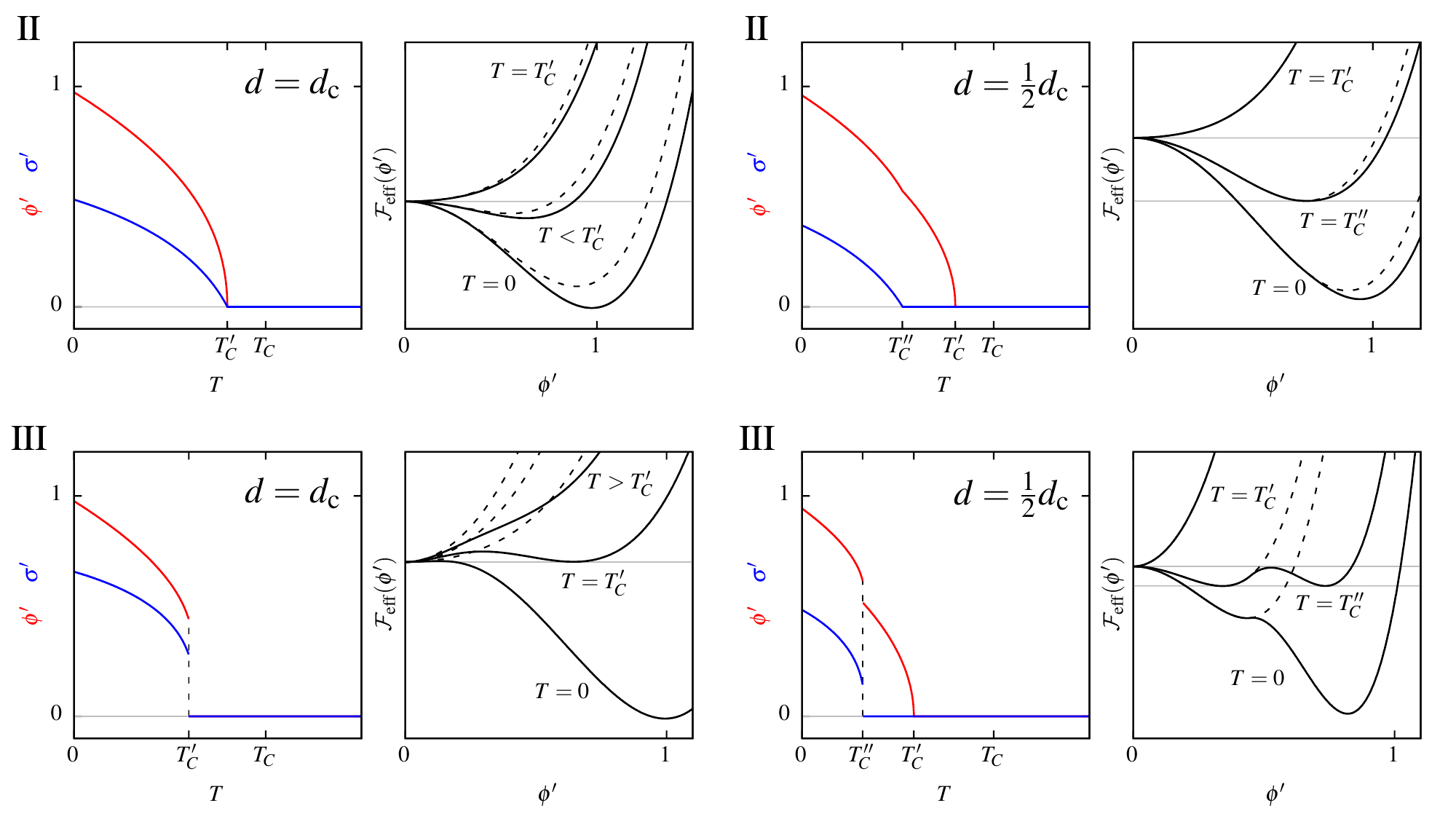}
\caption{Summary of the carrier transitions with temperature for the two coupled regions in Fig.~\ref{fig:ch-3-phase-diagram-temp}. The left hand side shows transitions for thin films at the uncoupled critical thickness ($d=\dc$) and the right hand side shows transitions for sub-critical thin films ($d=\frac{1}{2}\dc$). The right panels shows energy contours for different temperatures and are as described in Fig.~\ref{fig:ch-3-regions-thickness}, but an effective theory in $\p'$ instead of $\s'$ is used.}
\label{fig:ch-3-regions-temp}
\end{figure*}

In the previous section, the positions of the extrema were not dependent on the quantity that induced the transitions. Assuming the transition occurs at some temperature $T^*$, we can determine the character of the transition by requiring that $f_T(\s_{\text{min}}(T^*)) = 0$,
\beq{}
\begin{split}
-\Lambda \lb \sqrt{\frac{\Gamma(T^*)}{3\Lambda}}\rb^3 &+ \Gamma(T^*)\lb \sqrt{\frac{\Gamma(T^*)}{3\Lambda}} \rb \lb\frac{P_s^2/\ep}{A}\rb\frac{\dc}{d} = 0\\
\implies & \Gamma(T^*) = \sqrt[3]{\frac{27}{4}\lb \lb\frac{P_s^2/\ep}{A}\rb\frac{d_c}{d}\rb^2\Lambda}
\end{split}
\eec
and also requiring that $\s_{\text{min}}(T^*) > 0$,
\beq{}
\frac{1}{4}\lb\frac{P_s^2/\ep}{A}\rb\frac{\dc}{d} < \Lambda
\eep
Thus, we get
\beq{eq:ch-3-curved-line-temp}
A > \sqrt{\frac{1}{4}\frac{\dc}{d}\lb \frac{P_s^2}{\ep}\rb\Xpi}
\eeq
The lines in Eqs.~\eqref{eq:ch-3-straight-line-temp} and \eqref{eq:ch-3-curved-line-temp} form a diagram for transitions with temperature, shown in Fig.~\ref{fig:ch-3-phase-diagram-temp}. There are only two distinct regions (II and III), but the number of transitions in each is determined by the ratio $\frac{d}{\dc}$, so four different types of carrier transitions are possible. Plots of the order parameters for the four different scenarios are shown in Fig.~\ref{fig:ch-3-regions-temp}.

\subsection{Transitions with applied field}

\begin{figure}[ht] 
\centering
\includegraphics[width=0.9\linewidth]{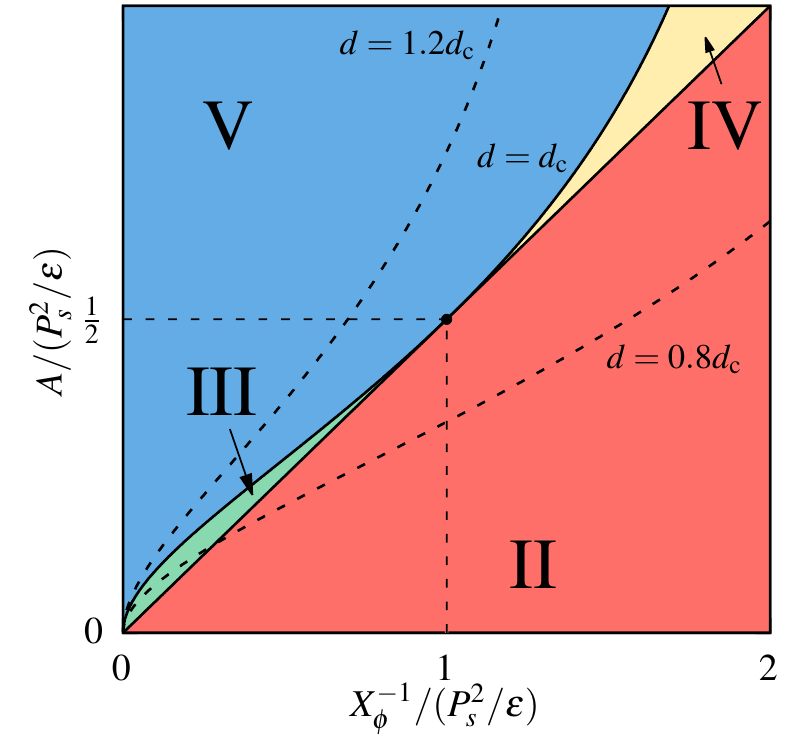}
\caption{Diagram for transitions with applied field. The curved line depends on the ratio $\frac{d}{\dc}$. The dashed black line is $\frac{d}{\dc} = 1$, which touches the straight line at $(A^*,X_{\p}^{-1*})$. Curves for $\frac{d}{\dc}=0.8,1.2$ are also sketched. First-order transitions should be observed below the curved line but above the straight line, i.e.~regions III and IV.}
\label{fig:ch-3-diagram-field}
\end{figure}

\begin{figure}[ht!] 
\centering
\includegraphics[width=\linewidth]{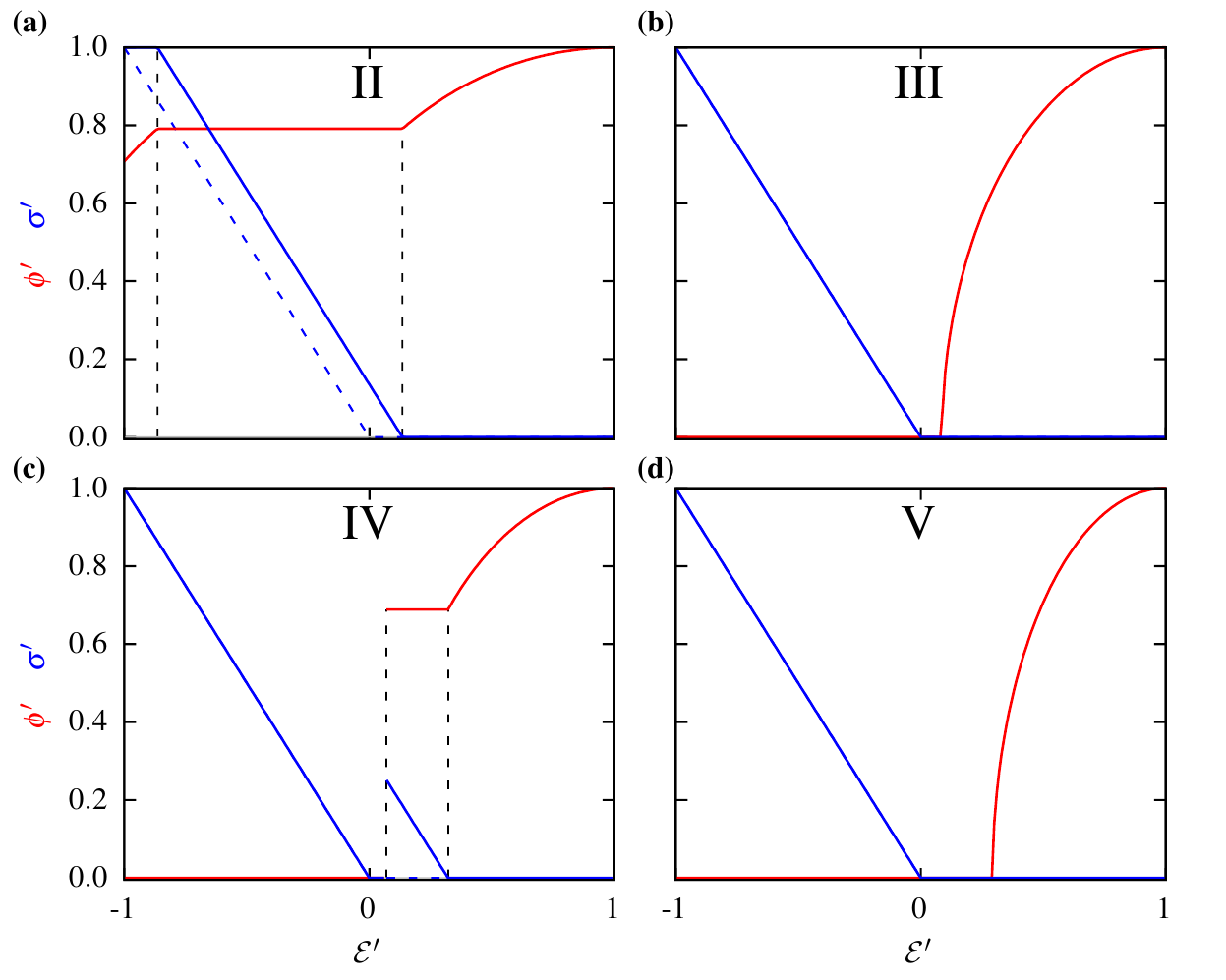}
\caption{Summary of the carrier transitions with applied applied field for the four coupled regions in Fig.~\ref{fig:ch-3-diagram-field}. The dashed blue line shows the value of $\s'$ without coupling to tilts. All plots are for $\frac{d}{\dc}=1$, and the following coorindates in Fig.~\ref{fig:ch-3-diagram-field}: \textbf{(a)}: $\Xpi$ = 1, $A$ = 0.25, \textbf{(b)}: $\Xpi$ = 0.25, $A$ = 0.15, \textbf{(c)}: $\Xpi$ = 1.75, $A$ = 1, \textbf{(d)}: $\Xpi$ = 1, $A$ = 1.}
\label{fig:ch-3-regions-nonzero-field}
\end{figure}

For transitions with applied field, note from the left-hand side of Eq.~\eqref{eq:ch-3-sigma-condition} that if $\s=0$, there is a critical field value at which the tilts switch on or off:
\beq{}
\Ef_{\p} = 1- \frac{1}{\sqrt{2A\Xp}} \geq 0
\eeq
We get $\Ef_{\p}=0$ when $A=\frac{1}{2}\Xpi$, which is the same condition for transitions with thickness at zero field. In order to determine whether transitions with applied field are continuous or discontinuous, we must solve Eq.~\eqref{eq:ch-3-cubic}, assuming there is some critical field at which carriers appear, i.e.~we must solve $f(\s'_{\text{min}}(\Ef'^*))=0$ for $\Ef'^*$ in a similar manner to the previous section. The boundary between continuous and discontinuous transitions of the tilts with applied field is:
\beq{eq:curve_E}
\Xpi = \frac{27}{2} \frac{\lb \dc/d\rb^2}{\lb 1+\frac{P_s^2/\ep}{A}\rb^3}\frac{1}{A}
\eec
which in general depends on the ratio $\frac{d}{\dc}$. For $\frac{d}{\dc}=1$, Eq.~\eqref{eq:curve_E} touches the straight line at the point $(A^*, X_{\p}^{-1 *})$. When $\frac{d}{\dc} < 1$, the two lines intersect before $(A^*,X_{\p}^{-1*})$. When $\frac{d}{\dc} > 1$, the two lines never intersect. The diagram describing the different possible transitions with applied field is sketched in Fig.~\ref{fig:ch-3-diagram-field} for $\frac{d}{\dc}=1$, and plots of the order parameters for each region are shown in Fig.~\ref{fig:ch-3-regions-nonzero-field}. In all except region II, the tilts are zero at zero field. When a negative field is applied, the carriers appear as in the uncoupled model. In region II the carriers grow linearly with a negative applied field, and can be switched off with a positive applied field. In all regions, tilts appear when a positive field is applied, because a positive field suppresses the polar mode. Although first-order transitions of the tilts should be observed in regions III and IV, we did not find any first-order transitions for physically meaningful values of the order parameters or applied field in region III. A first-order transition of the tilts and carriers for a positive applied field was observed in region IV, however.

\section{Coupling to inhomogeneous tilts}

\begin{figure}[ht] 
\centering
\includegraphics[width=\columnwidth]{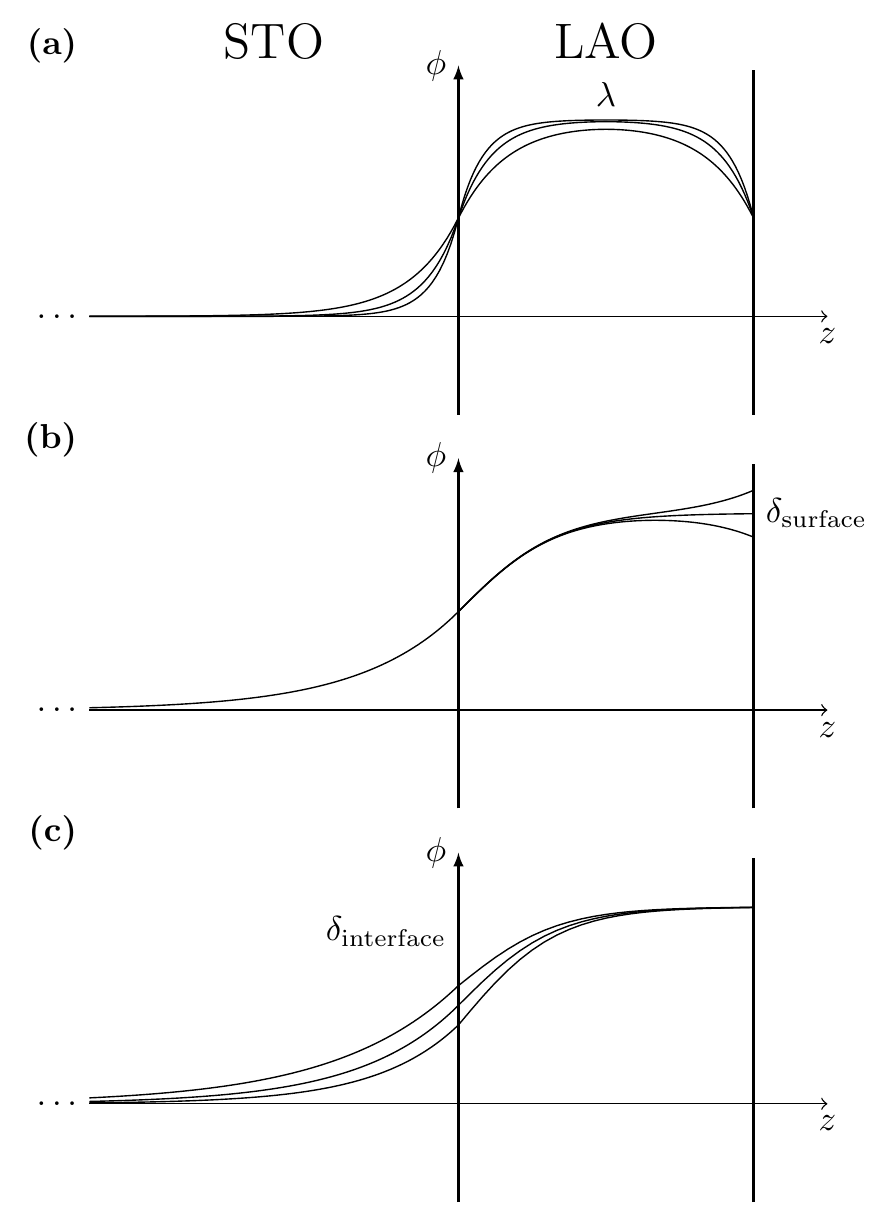}
\caption{Sketch of inhomogeneous tilts in a thin film grown on a substrate, e.g.~LAO/STO, and the effect of \textbf{(a)}: $\lambda$, the correlation length of the tilts in the LAO film \textbf{(b)}: $\d_{\text{surface}}$, the relative difference in the energy of the tilts at the surface with respect to the bulk and \textbf{(c)}: $\d_{\text{interface}}$, the relative difference in the energy of the tilts at the interface with respect to the bulk. The effect of the energetics of the tilts in the substrate may be absorbed into $\d_{\text{interface}}$.}
\label{fig:ch-4-diagram-GL}
\end{figure}

We can allow for inhomogeneous tilts by generalizing from a Landau theory to a Ginzburg-Landau theory, i.e.~the free energy is expanded in powers of both the order parameter and its gradient. We do this following the methodology which has been used to describe semi-infinite ferromagnetic \cite{lubensky1975critical}, ferroelectric \cite{kretschmer1979surface,chandra2007landau,PhysRevLett.102.147601} and superconducting \cite{de2018superconductivity} systems in which the order parameter is inhomogeneous due to the presence of a free surface. Considering only the tilts in the thin film, Eq.~\eqref{eq:ch-3-F-tilt-1} becomes an integral over the volume of the film:
\beq{eq:ch-4-F-tilt-GL}
{\F'}_{\p}^{\text{GL}} = \frac{1}{V}\int_{V} \left[\la^2\lb\grad \p'(\bm{r})\rb^2 + \frac{1}{4}\p'(\bm{r})^4 -\frac{1}{2}\p'(\bm{r})^2 \right]\dd{V}
\eec
where ${\F'} \equiv \F/\frac{1}{2}\Xpi$ and $\la > 0$ is the correlation length of the tilts, which sets the length scale for the variations of the tilts, see Fig.~\ref{fig:ch-4-diagram-GL} (a). The correlation length depends in general on the temperature of the system, but here we assume a fixed value for simplicity. Eq.~\eqref{eq:ch-4-F-tilt-GL} is written so that each term in the integrand is dimensionless, and the free energy is in units of $\frac{1}{2}\Xpi$, the energy scale of the homogeneous tilts.

Next, we reintroduce the carriers. We take the order parameter to be the carrier concentration averaged over an in-plane unit cell: $\s'\equiv \left<\s'\right> = \frac{1}{A}\int \s'(x,y)\dd{A}$. If we assume that the in-plane variance in $\s'$ is small, so that $\left<(1-\s')\right>^2 \approx (1-\left<\s'\right>)^2$, then ${\F'}_{\s}^{\text{GL}} = {\F'}_{\s}$. It should be noted that this is a significant assumption, especially in the low carrier regime. 

For inhomogeneous tilts, the biquadratic coupling term becomes
\beq{eq:ch-4-F_couple_GL}
{\F'}_{\s\p}^{\text{GL}} = \frac{1}{V}\int \frac{1}{2} (2A\Xp) \p'(\textbf{r})^2 (1-\s')^2 \dd{V}
\eec
which, since we assume the in-plane variance in $\s'$ is small, reduces to
\beq{}
{\F'}^{\text{GL}}_{\s\p}  = \frac{1}{2} (2A\Xp) (1-\s')^2 \left< \p'^2\right>
\eec
where 
\beq{}
\left< \p'^2\right> = \frac{1}{V}\int \p'(\textbf{r})^2 \dd{V}
\eeq
is the mean of the square of the tilt throughout the film. 

Under these assumptions, the free energy can be minimized with respect to $\s'$ as in the homogeneous case, yielding
\beq{eq:ch-4-GL-sigma}
\s' = 1 - \frac{1}{1+\frac{A}{P_0^2/\ep}\left< \p'^2\right>}\lb \frac{\dc}{d} + \Ef'\rb
\eec
which is the same as the second line of Eq.~\eqref{eq:ch-3-solutions-reduced}, but with ${\p'^2\to\left<\p'^2\right>}$. We only consider transitions with thickness for illustrative purposes, although it would be simple enough to include an applied field as we did in the previous section. 

For the tilts, we have
\beq{}
\begin{split}
{\F'}_{\p}^{\text{GL}} + {\F'}_{\s\p}^{\text{GL}} = &\frac{1}{V}\int_{V} \left[\la^2\lb\grad \p'\rb^2 + \frac{1}{4}\p'^4\right. \\
&\left. \qquad\quad -\frac{1}{2}\lb 1-2AX_{\p}(1-\s')^2\rb\p'^2 \right]\dd{V}
\end{split}
\eep
Following a similar treatment of ferroelectric thin films with inhomogeneous polarization \cite{chandra2007landau}, we split the free energy into interior and surface contributions: ${\F^{\text{GL}} = \F^{\text{GL}}_{\text{interior}}+\F^{\text{GL}}_{\text{boundary}}}$, where $\F^{\text{GL}}_{\p,\text{boundary}}$ is given in terms of an expansion of a local order parameter at the boundaries. First, we use the reverse product rule to rewrite the gradient term:
\beq{eq:ch-4-stokes}
(\grad\p')^2 = \grad\cdot(\p'\grad\p') - \p'\grad^2\p'
\eec
then we use Stokes' theorem on the first term in Eq.~\eqref{eq:ch-4-stokes}:
\beq{}
\begin{split}
\int (\grad\p')^2 \dd V &= \int \left[\grad\cdot(\p'\grad\p') - \p'\grad^2\p' \right]\dd V\\
& = \int \p'\grad\p'  \dd S- \int  \p'\grad^2\p' \dd V
\end{split}
\eec
where for a thin film on a substrate the surface integral has two terms: one at the free surface and one at the interface with the substrate:
\beq{eq:ch-4-F-GL}
\begin{gathered}
{\F'}^{\text{GL}} = \frac{1}{V}\int_{V} \bigg[\frac{1}{4}\p'^4 -\frac{1}{2}\lb 1-2A\Xp(1-\s')^2\rb\p'^2\\
- \la^2\p'\grad^2\p'\bigg]\dd{V}\\
+ \frac{1}{V}\int_{\text{s}}\left[\la^2\lb\hat{n}\cdot\grad\p'\rb\p'\right]\dd{S}\\
 + \frac{1}{V}\int_{\text{i}}\left[\la^2\lb\hat{n}\cdot\grad\p'\rb\p'\right]\dd{S}
\end{gathered}
\eec
where s and i refer to the surface and interface, respectively. Note that the surface free energies are also in units of $\frac{1}{2}X_{\p}^{-1}$, and the surface integrals are dimensionless.

Next we claim that there must be a difference in the energies at the boundaries compared to the interior of the film \cite{lubensky1975critical,de2018superconductivity,chandra2007landau}, which can be described using an expansion of the local order parameters at the boundaries:
\beq{}
\begin{split}
{\F'}^{\text{GL}}_{\p,\text{boundary}} &=\frac{1}{V}\int_{\text{s}}\left[\la^2\lb\hat{n}\cdot\grad\p'\rb\p' + \frac{1}{2}\d_{\text{s}}\p'^2\right]\dd{S}\\ 
&+ \frac{1}{V}\int_{\text{i}}\left[\la^2\lb\hat{n}\cdot\grad\p'\rb\p' + \frac{1}{2}\d_{\text{i}}\p'^2 \right]\dd{S}
\end{split}
\eep
The quadratic coefficient of the surface order parameter describes the relative difference between the energy of the tilts at the surface and the energy per unit volume of the tilts in the bulk, and therefore must have units of length. We call this the extrapolation length, $\d$. $\d_{\text{s}}$ and $\d_{\text{i}}$ are the surface and interface extrapolation lengths, respectively, which describe the difference in energy at the surface and interface with respect to the bulk, see Figs.~\ref{fig:ch-4-diagram-GL} (b) and (c). In general, higher order terms could be included in this expansion, but we assume that the leading term is the quadratic one \cite{chandra2007landau}. Also, the boundaries may have a temperature dependence which differs from the bulk, which could lead to a change in sign of the extrapolation lengths.

Minimizing the total free energy in the bulk and at both boundaries, we get
\beq{eq:ch-4-GL-equation}
\begin{gathered}
\p'^3 - \lb 1-2A\Xp(1-\s')^2\rb\p' - \la^2\grad^2 \p' = 0 \\
(\hat{n}\cdot(\grad\p')) +\frac{\d_{\text{i}}}{\la^2}\p' = 0, \qquad z = 0 \\
(\hat{n}\cdot(\grad\p')) +\frac{\d_{\text{s}}}{\la^2}\p' = 0, \qquad z = d \\
\s' = 1 - \frac{1}{1+\frac{A}{P_0^2/\ep}\left< \p'^2\right>}\lb \frac{\dc}{d} + \Ef'\rb
\end{gathered}
\eep
This is the generalization of Eq.~\eqref{eq:ch-3-solutions-reduced} but with inhomogeneous tilts. The expression for $\s'$ is almost identical, but with the square of the tilt replaced with the mean of the square of the tilt throughout the film. $\p'$ is described by a second order differential equation plus two boundary conditions, one for the free surface and one for the interface. These boundary conditions are referred to as Robin boundary conditions, i.e.~a linear combination of the tilt and its gradient at each boundary. The equations for $\p'$ and $\s'$ are again self-consistent, making them difficult to solve. In the previous section we used physical constraints to determine the character of the carrier and tilt transitions under various conditions, summarized in Fig.~\ref{fig:ch-3-phase-diagram}, but this would be difficult or impossible for Eq.~\eqref{eq:ch-4-GL-equation}. Thus, our only option is obtaining numerical solutions, solving for $\s'$ and $\p'$ iteratively until both are converged below a suitable tolerance. This approach was used in the previous section to obtain the order parameter plots shown in Figs.~\ref{fig:ch-3-regions-thickness}, \ref{fig:ch-3-regions-temp} and \ref{fig:ch-3-regions-nonzero-field}. 

Eq.~\eqref{eq:ch-4-GL-equation} is solved as follows: the tilt profile inside the film is obtained by solving the first three lines using finite difference methods. The mean of the tilt squared in the film is then used to calculate the carrier concentration, which is re-inserted into the ODE. The two are solved self-consistently until their relative changes between successive iterations are below a suitable tolerance. 

For a given set of system parameters, we can obtain order parameter plots as a function of thickness or applied field, similar to Figs.~\ref{fig:ch-3-regions-thickness}, \ref{fig:ch-3-regions-temp} and \ref{fig:ch-3-regions-nonzero-field}. The main difference with the inhomogeneous tilts is that $\la$, $\d_{\text{s}}$ and $\d_{\text{i}}$ need to be specified, in addition to $\Xpi$, $A$ and $\dc$. Thus, for inhomogeneous tilts, Fig.~\ref{fig:ch-3-phase-diagram} becomes five dimensional, and systemically navigating such a diagram would be impractical.

It is also difficult to estimate $\la$, $\d_{\text{s}}$ and $\d_{\text{i}}$. Previous studies have fixed $\la$ to a small number of unit cells \cite{kretschmer1979surface}. In Ref.~\onlinecite{stengel_tilt}, $\la$ is estimated to be around 1-2 unit cells in bulk LAO from calculations of the dispersion of the phonon branch corresponding to the $a^{-}a^{-}c^0$ tilt. To our knowledge, it is not possible to directly measure the extrapolation lengths. In principle, one could fit measurements of tilts in a thin film, either from experiment or first-principles calculations, to Eq.~\eqref{eq:ch-4-GL-equation}, but the tilts at the boundaries are strongly affected by both the correlation length and the extrapolation length, which would lead to overfitting.

It is helpful to note that, since we only consider indirect coupling, the carrier concentration depends only on the mean of the square of the tilts, and not on the specific shape of the tilts. Thus, we can understand the influence of the inhomogeneity of the tilts by fixing the extrapolation lengths and allowing the correlation length to change.

\begin{figure}[ht] 
\centering
\includegraphics[width=\columnwidth]{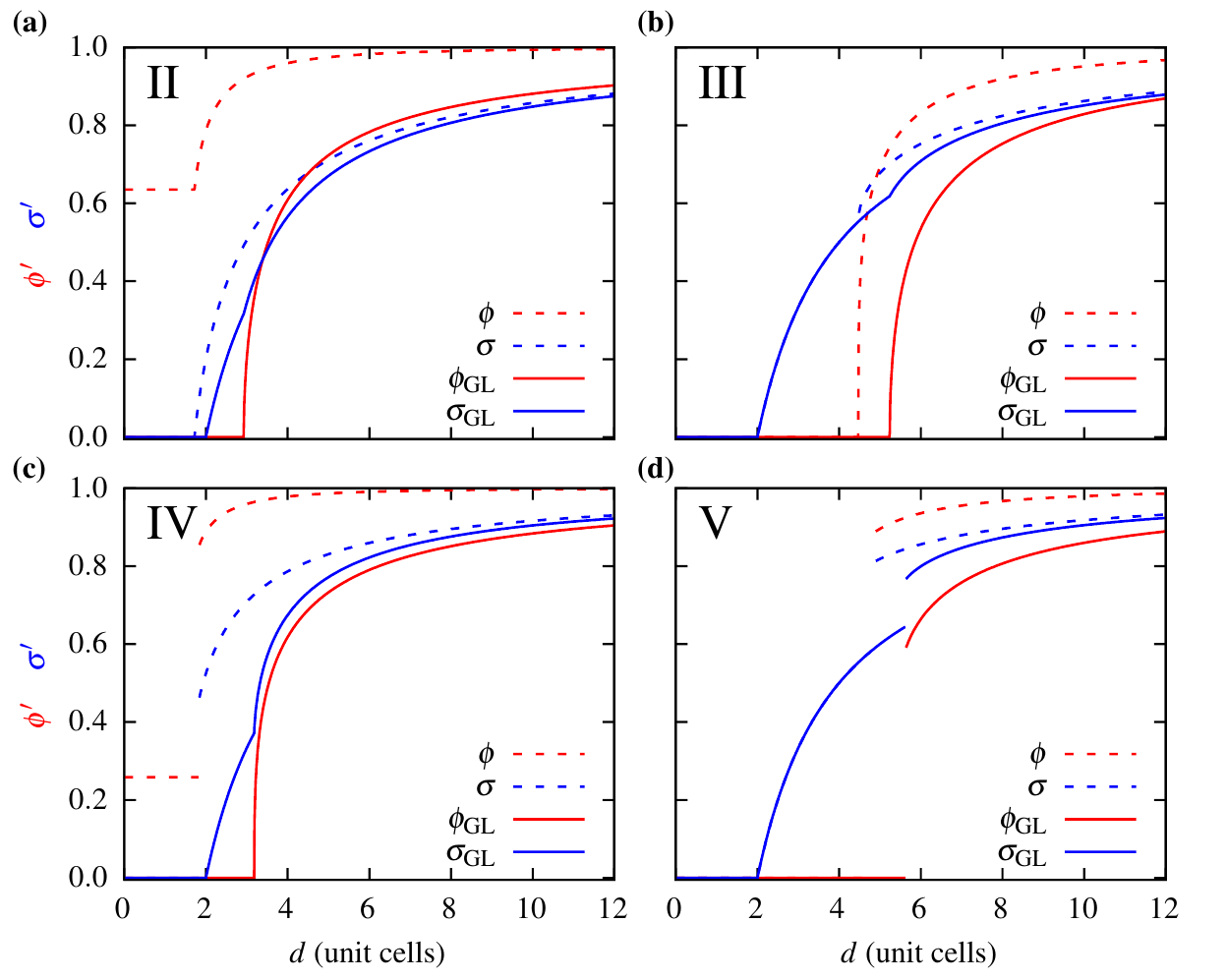}
\caption{Order parameter plots in the four coupled regions in Fig.~\ref{fig:ch-3-phase-diagram}, for both homogeneous (dashed lines) and inhomogeneous (solid lines) tilts. \textbf{(a)}: Region II (red): $\Xpi = 1.36454$, $A = 0.407$. \textbf{(b)}: Region III (green): $\Xpi = 0.2$, $A = 0.5$. $\s$ and $\s_{\text{GL}}$ are identical between $\dc$ and $\dphi$. \textbf{(c)}: Region IV (yellow): $\Xpi = 3$, $A = 1.4$. \textbf{(d)}: Region V (blue): $\Xpi = 0.5$, $A = 1.5$. $\s$ and $\s_{\text{GL}}$ are identical between $\dc$ and $\dphi$. In all regions, $\dc = 2$ unit cells.}
\label{fig:ch-4-GL-all}
\end{figure}

First, we examine the behavior of the order parameters in the different regions in Fig.~\ref{fig:ch-3-phase-diagram} and compare with the results obtained for homogeneous tilting. For each region, we set $\la=2$ unit cells, $\dc = 2$ unit cells, the thickness at which a density of trapped Ti 3$d$-like states has been observed in LAO/STO \cite{sing2009profiling,berner2010laalo,
takizawa2011electronic,slooten2013hard}, $\d_{\text{i}} = 5$ unit cells and $\d_{\text{s}} = 0$. The order parameter plots are shown in \ref{fig:ch-4-GL-all} (solid lines), alongside the corresponding result from the homogeneous theory (dashed lines). We see that, for this choice of parameters, all of the points except the one in region V shift to region III when the tilts become inhomogeneous. In all cases, the carriers appear continuously at $\dc$, and $\dphi$ increases with respect to the corresponding scenario from the homogeneous theory. This makes sense, since at smaller thicknesses there would be a large gradient in the tilts and hence a large energy penalty. Thus, the tilts can only appear at larger thicknesses, where the magnitude is roughly constant inside the bulk of the film and only changes in a small region near both boundaries.

\begin{figure}[ht] 
\centering
\includegraphics[width=\columnwidth]{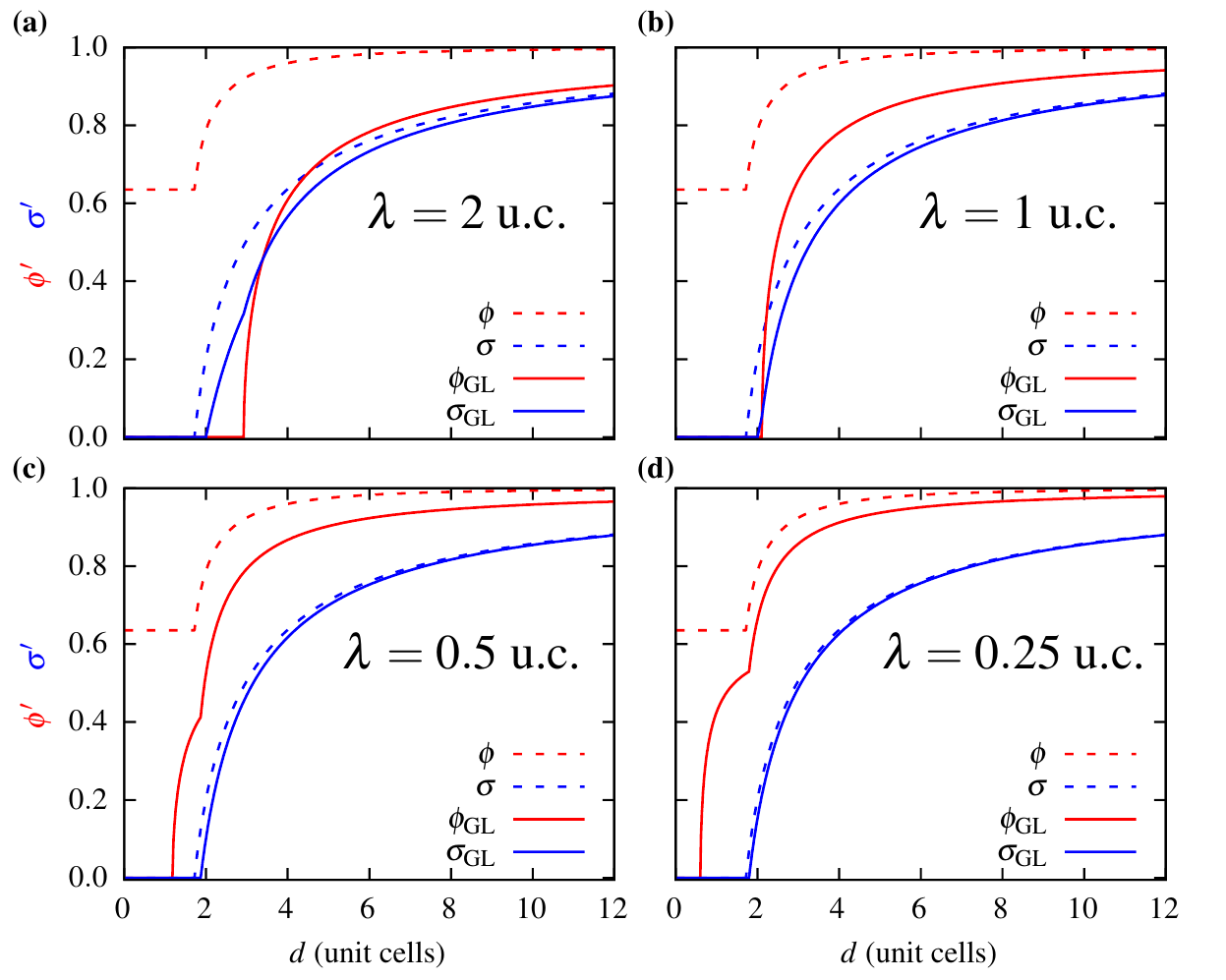}
\caption{Order parameter plots in region II for several values of $\la$, for both homogeneous (dashed lines) and inhomogeneous (solid lines) tilts. \textbf{(a)}: $\la = 2$ unit cells, \textbf{(b)}: $\la = 1$ unit cell, \textbf{(c)}: $\la = 0.5$ unit cells, and \textbf{(d)}: $\la = 0.5$ unit cells.}
\label{fig:ch-4-GL-II-all}
\end{figure}

\begin{figure}[ht] 
\centering
\includegraphics[width=\columnwidth]{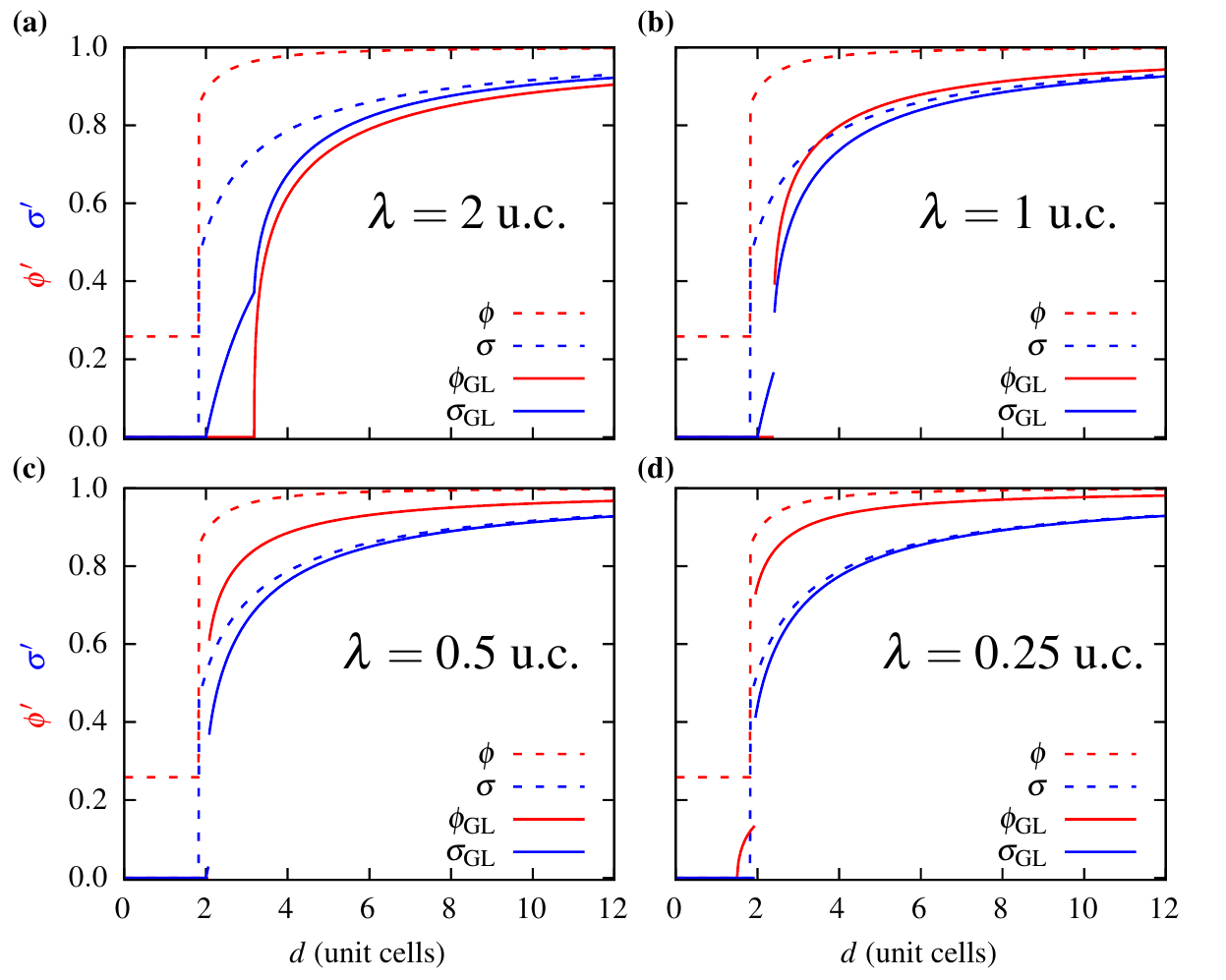}
\caption{Order parameter plots in region IV for several values of $\la$, for both homogeneous (dashed lines) and inhomogeneous (solid lines) tilts. \textbf{(a)}: $\la = 2$ unit cells, \textbf{(b)}: $\la = 1$ unit cell, \textbf{(c)}: $\la = 0.5$ unit cells, and \textbf{(d)}: $\la = 0.5$ unit cells.}
\label{fig:ch-4-GL-IV-all}
\end{figure}

Next, we examine the effect of changing $\la$. We found that for regions III and V, the transitions remain in regions III and V, but the value of $\dphi$ is sensitive to $\la$. For regions II and IV, the points move through different regions as $\la$ decreases, shown in Figs.~\ref{fig:ch-4-GL-II-all} and \ref{fig:ch-4-GL-IV-all}, respectively. As $\la$ decreases, and there is less of a penalty for the inhomogeneity of the tilts, $\dphi$ decreases, and eventually the order of the transitions reverse, which occurs continuously in region II (Figs.~\ref{fig:ch-4-GL-II-all} (c) and (d)) and discontinuously in region IV (Fig.~\ref{fig:ch-4-GL-IV-all} (d)). These types of transitions do not correspond to any of regions in Fig.~\ref{fig:ch-3-phase-diagram}, so we label them regions VI and VII. In region VI, tilts first appear continuously at $\dphi$, and then carriers at a critical thickness between $\dphi$ and $\dc$. In region VII, the order of the transitions with thickness is the same, but the carriers appear discontinuously.

\begin{figure}[ht] 
\centering
\includegraphics[width=\columnwidth]{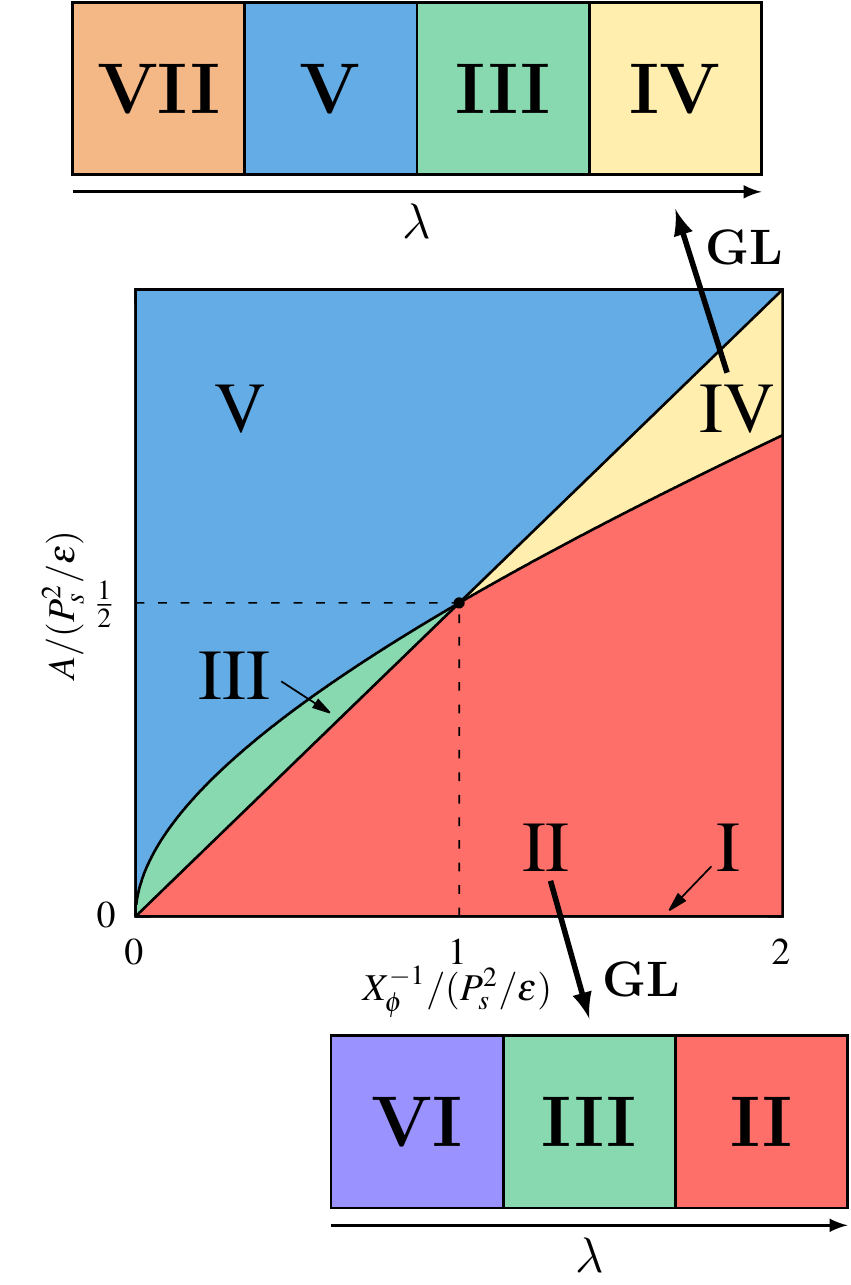}
\caption{Summary of the additional carrier and tilt transitions which can occur when the tilts are inhomogeneous. The sequence of transitions which can occur in regions II and IV by changing $\la$ (fixing the rest of the parameters) are shown below and above the diagram, respectively.}
\label{fig:ch-4-phase-diagram-GL}
\end{figure}

In Fig.~\ref{fig:ch-4-phase-diagram-GL} we summarize the transitions observed in the inhomogeneous theory. Regions III and V remain unchanged, with the exception that $\dphi$ is renormalized by $\la$. For regions II and IV, the type of transition can change, depending on the value of $\la$ (and the other parameters). The different transitions that can occur by changing $\la$ but fixing the rest of the parameters are shown. For $\la\to\infty$, the homogeneous case is realized. As $\la$ decreases, the character of the transition changes. For sufficiently small $\la$, two entirely new types of transitions occur, which we label region VI (Figs.~\ref{fig:ch-4-GL-II-all} (c) and (d)) and region VII (Fig.~\ref{fig:ch-4-GL-IV-all} (d)).

\section{Discussion and Conclusion}\label{section:conclusions}

In this paper, we introduced a phenomenological theory which considers the influence of homogeneous tilts on the appearance of carriers at polar-nonpolar perovskite interfaces. We show that, upon coupling, four new types of transitions of tilts and carriers with film thickness or applied field are possible, depending only on the energetics of the tilts, the polar discontinuity, and the biquadratic coupling between the tilts and the polar mode. These include the simultaneous appearance of tilts and carriers at a reduced thickness compared to the uncoupled theory, which can be either continuous or discontinuous, and separate transitions, with carriers appearing first and then tilts, the second transition being either continuous or discontinuous.

First-principles calculations of bulk LAO were performed to predict the character of the transitions at the LAO/STO interface. For various exchange-correlation functionals, all of the calculations predicted that at zero temperature and for homogeneous tilts, LAO/STO lies in region II of Fig.~\ref{fig:ch-3-phase-diagram}, i.e.~a single continuous transition of both tilts and carriers. We also investigated the possibility of tuning the character of the transition via biaxial strain by changing the in-plane lattice constant by up to $\pm$1\% and repeating the calculations. We found that the position of the point in Fig.~\ref{fig:ch-3-phase-diagram} can be changed by adjusting the strain, but not by enough to change the character of the transitions. However, in general, if a system were closer to a boundary between regions, it may be possible to change the regions using biaxial strain. Additionally, if compressive strain is applied to $a_{\text{LAO}}$, the tilt pattern changes to $a^0a^0c^-$ \cite{tilt_strain}, which would change the competition with the polar mode. This could be achieved by using a nonpolar perovskite with a smaller lattice constant than LAO as the substrate material.

In addition to changing the strain and the tilt pattern as mentioned above, changing the materials which form the polar-nonpolar interface can also change the chemistry at the interface, and hence the polar discontinuity. Different interfaces have been investigated experimentally \cite{temperature_experiment,kumar2020lasco3,
wang2016creating,moetakef2011electrostatic}, computationally \cite{plos_paper,maznichenko2019formation,kim2018laino3}, and high throughput searches for new interfaces have been done \cite{high_throughput}. Carrier transitions have also been observed at spinel-perovskite interfaces \cite{spinel_1,spinel_2,spinel_3}. Thus, it may be possible to observe a wide variety of behavior in the carrier transitions at different polar-nonpolar perovskite interfaces for different combinations of materials. More specifically, by estimating where different systems would lie on Fig.~\ref{fig:ch-3-phase-diagram}, it may be possible to establish an atlas of transitions across different possible polar-nonpolar perovskite interfaces.

We also showed how the appearance of carriers can be tuned via temperature through the temperature dependence of the tilts. At a finite temperature, the curvature of the tilt double well is ${\Xpi(T) = \Xpi\lb 1-\frac{T}{T_C}\rb}$, so increasing the temperature would move any point on the diagram in Fig.~\ref{fig:ch-3-phase-diagram} to the left. We would expect the dependence of the biquadratic coupling term $A(T)$ on temperature to be similar but weaker. Thus, it may be possible to move a system from one point in the diagram to another if the initial point is sufficiently close to a boundary. In Ref.~\onlinecite{fister2014octahedral}, when grown on a substrate of cubic STO, LAO was found to undergo a transition from untilted to tilted below $695^{\circ}\si{C}$ for a thickness of 24 unit cells and $540^{\circ}\si{C}$ for a thickness of 9 unit cells, where both samples were under the same strain conditions. Because both films were supercritical there was no carrier transition associated with the tilt transitions, but their observations are in agreement with our prediction of a second transition temperature, $T_C''$, which is sensitive to the ratio $\frac{d}{\dc}$. Assuming that the films were sufficiently thick so that the effect of gradients are negligible, this may provide evidence for the additional transition temperature arising from the coupling between tilts and carriers.

We also investigated the influence of the inhomogeneity of the tilts on the coupling between tilts and carriers. Upon coupling to inhomogeneous tilts, the diagram in Fig.~\ref{fig:ch-3-phase-diagram} becomes five dimensional, with additional axes for the correlation length of the tilts in the polar film, and extrapolation lengths for the interface and the free surface of the film. Practically, it would be very demanding to explore this five dimensional space numerically, and to our knowledge, it can't be done analytically. However, since the biqaudratic coupling between tilts and carriers only depends on the mean of the tilt squared throughout the film, we can examine the influence of inhomogeneous tilts by examining the effect of the correlation length on transitions in each region in Fig.~\ref{fig:ch-3-phase-diagram}, with the rest of the parameters fixed. We found that, for the regions corresponding to the separate appearances of tilts and then carriers, the order of the transitions is unaffected by the correlation length, except for the value of the second transition shifting slightly. For the regions corresponding to simultaneous transitions, we found that a finite correlation length can actually change the character of the transition, and can lead to two new types of transitions which are not possible for homogeneous tilts, where the tilts appear before carriers, either continuously or discontinuously.

Due to the large number of system parameters required, many of which cannot be directly measured, it remains difficult to conclusively determine the order of the carrier transition at polar-nonpolar perovskite interfaces. However, we have shown that, even using a phenomenological description, an extremely rich variety of behavior is possible for the appearance of carriers at polar-nonpolar perovskite interfaces. This includes first-order transitions, which have been observed experimentally \cite{ohtomo2004}, and have been claimed to occur in other theoretical studies \cite{zunger_dft_paper}, but without a detailed explanation.

\section*{Acknowledgments}
DB would like to acknowledge funding from the EPSRC Centre for Doctoral Training in Computational Methods for Materials Science under grant number EP/L015552/1, St.~John's College (University of Cambridge), and the University of Li{\'e}ge under special funds for research (IPD-STEMA fellowship programme). PAP would like to acknowledge funding from the Diputaci\'on Foral de Gipuzkoa through Grant 2020-FELL-000005-01.
  Funding from the Spanish MICIN is also acknowledged, 
through grant PID2019-107338RB- C61/AEI/10.13039/501100011033, 
as well as a Mar\'{\i}a de Maeztu award to Nanogune, Grant 
CEX2020-001038-M funded by MCIN/AEI/ 10.13039/501100011033.


\section*{Appendix A: First-Principles Calculations}

We can approximate $\Xpi$, $A$ and $\frac{P_s^2}{\ep}$ for LAO/STO by performing first-principles calculations using bulk LAO, biaxially strained to the lattice parameter of cubic STO. We follow the methodology in Ref.~\onlinecite{stengel_tilt}, where similar calculations were performed.

First-principles density functional theory (DFT) calculations were performed using the {\sc abinit} code \cite{abinit5,abinit4,abinit3,abinit2,abinit1}, version 8. We used both Perdew-Burke-Edwards (PBE) \cite{pbe} and PBEsol \cite{pbesol,libxc} exchange-correlation functionals within the generalized gradient approximation (GGA) using {\sc psml} \cite{psml} norm-conserving \cite{norm_conserving} pseudopotentials, obtained from pseudo-dojo \cite{pseudodojo}. We also performed calculations using the Perdew-Wang (PW92) \cite{pw92} exchange-correlation functional within the local density approximation (LDA), using the projector augmented-wave method \cite{paw,torrent2008paw} (PAW) in order to compare to Ref.~\onlinecite{stengel_tilt}. For the electronic configurations, we included 11 valence electrons for La (\chem{5s^2 5p^6 5d^1 6s^2}), 11 for Al (\chem{2s^2 2p^6 3s^2 3p^1}) and 6 for O (\chem{2s^2 2p^4}), explicitly including the semicore \chem{2s^2 2p^6} states in the valence configuration of Al. When the semicore electrons were included for Al, a cutoff of $2500 \ \si{eV}$ was required to adequately converge the total energy of the 5-atom unit cell. A Monkhorst-Pack $k$-point grid \cite{mp} of $6 \times 6 \times 6$ was used for the 5-atom calculations and a grid of $4\times 4\times 3$ was used for the 20-atom calculations.

\begin{table*}[ht!]
\renewcommand*{\arraystretch}{1.5}
\setlength{\tabcolsep}{7pt}
\begin{center}
\begin{tabular}{c | c c c | c c | c c c c }
\hline\hline
Functional & $a_{\text{LAO}}$ (\AA) & $a_{\text{STO}}$ (\AA) & $c$ (\AA) & $E_0$ (meV/$\Omega_5$) & $\Xpi$ (meV/$\Omega_5$) & $\ep_5$ & $\ep_{20}$ & $\frac{P_s^2}{\ep_5}$ (meV/$\Omega_5$) & $A$ (meV/$\Omega_5$) \\ \hline
PAW LDA \cite{stengel_tilt} & 3.75 & 3.85 & 3.68 & 27.7 & 221.6 & - & - & - & 96.1 \\ \hline
PAW LDA & 3.75 & 3.86 & 3.67 & 35.9 & 287.8 & 26.1 & - & 425.7 & - \\ \hline 
PBE & 3.81 & 3.93 & 3.72 & 60.5 & 483.9 & 40.4 & 26.5 & 268.9 & 141.2 \\ \hline
PBEsol & 3.77 & 3.89 & 3.69 & 54.2 & 433.8 &  34.6 & 24.6 & 317.9 & 129.4 \\ \hline \hline
\end{tabular}
\end{center}
\caption{Summary of results from first-principles calculations. }
\label{table:ch-3-dft}
\end{table*}

We first optimized the geometry of cubic STO in order to obtain the lattice parameter, $a_{\text{STO}}$. The in-plane lattice parameters of LAO were then fixed to $a_{\text{STO}}$, and the out-of-plane lattice parameter $c$ was allowed to relax. It was found that the lattice parameter was not affected much more by the tilt or the electric fields applied, so it was fixed to the value obtained in the untilted case in the more time-consuming calculations. The untilted calculations were performed using the 5-atom primitive cell. To allow for tilts, we used a $\sqrt{2}\times\sqrt{2}\times 2$ supercell containing 20 atoms, which is the smallest cell required to allow for the $a^-a^-c^0$ tilt pattern which appears in LAO when biaxially strained to $a_{\text{STO}}$ \cite{tilt_strain}. The depth of the tilt double well, $E_0$, is the energy difference per 5 atom unit cell between the tilted and untilted systems. Using $\F_{\p}(\p' = \pm 1) = E_0$, we get
\beq{}
\Xpi = -8E_0
\eep
$E_0$ and $\Xpi$ are given in Table I for all of the functionals used, in units of $\si{meV/\Omega_5}$, where $\Omega_5$ is the volume of the 5-atom unit cell. The LDA results are close to those obtained in Ref.~\onlinecite{stengel_tilt}, but the results obtained using PBE and PBEsol differ by a factor of $\sim 2$.

\begin{figure}[ht!] 
\centering
\includegraphics[width=\columnwidth]{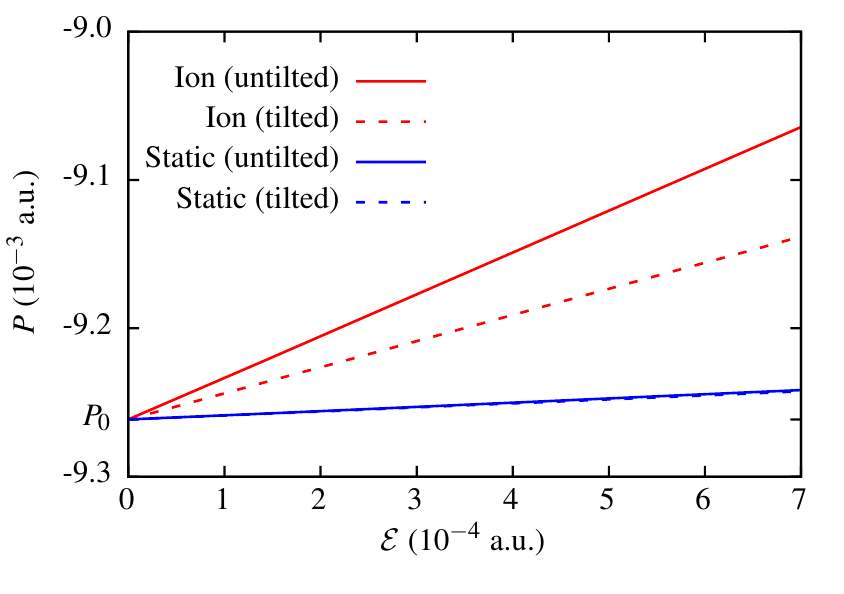}
\caption{Polarization as a function of applied electric field, from the PBEsol calculations. The solid lines are results obtained from the untilted system and the dashed lines are results obtained for the tilted system. The blue lines indicate the static case, where the ions were fixed and the red lines indicate the relaxed case, where a geometry relaxation was performed at each value of $\Ef$.}
\label{fig:ch-3-pbar}
\end{figure}

$A$ and $\frac{P_s^2}{\ep}$ were obtained by calculating the dielectric constants in the untilted and tilted systems. This was done by performing a set of bulk calculations using a finite electric field parallel to the $c$-axis \cite{PhysRevLett.73.712,PhysRevB.63.155107,PhysRevLett.89.117602,PhysRevLett.89.157602,PhysRevLett.96.056401}. In Ref.~\onlinecite{stengel_tilt} a finite displacement field was used \cite{stengel2009electric}, but either type of field can be used to calculate $A$. The polarization, calculated using Berry phases \cite{berry1984quantal,modern_theory_polarisation_1}, was measured in the presence of a small applied field $\Ef$, with the ions both fixed and allowed to relax. Plots of polarization as a function of applied field are shown in Fig.~\ref{fig:ch-3-pbar}. In Gaussian units, the relation between polarization and electric field, and dielectric constant is
\beq{}
\begin{split}
P_i &= \chi_{ij}\Ef_j + \bigO(\Ef^2)\\
\ep_{ij} &= 1+4\pi \chi_{ij}
\end{split}
\eec
The slope of $P(\Ef)$ is reduced in the tilted case, because the tilts compete with the polar mode. For the static calculations, the tilted and untilted systems gave identical results. When the tilts are allowed to relax, the relation between $P$ and $\Ef$ is
\beq{}
P(\Ef, \phi(\Ef)) = \lb \frac{1}{\ep_5}+\frac{A}{P_s^2}\p'(\Ef)^2\rb^{-1} \Ef
\eec
where $\ep_5$ is the dielectric constant of the untilted system. The dielectric constant of the tilted system, $\ep_{20}$, is obtained in the limit $\Ef\to 0$:
\beq{tilted_dielectric_2}
\frac{1}{\ep_{20}} = \frac{1}{\ep_5}+\frac{A}{P_s^2}
\eec
where we used $\p'(\Ef\to0) = 1$. Rearranging Eq. \eqref{tilted_dielectric_2} gives
\beq{}
\frac{\ep_{20}^{-1}-\ep_5^{-1}}{\ep_5^{-1}} = \frac{A}{(P_s^2/\ep_5)}
\eec
which is exactly the vertical axis of the phase transition diagram in Fig.~\ref{fig:ch-3-phase-diagram}. Values of $\ep_5$, $\ep_{20}$, $A$ and $\frac{P_s^2}{\ep_5}$ are given in Table~\ref{table:ch-3-dft} for all of the functionals used. For the LDA (PAW) calculations, the electric field calculations for the tilted system failed to converge, so we used the value of $A$ from Ref.~\onlinecite{stengel_tilt} in Fig.~\ref{fig:ch-3-phase-diagram}.\\

The results in Table~\ref{table:ch-3-dft} were used to place the LAO/STO interface on the phase transition diagram in Fig.~\ref{fig:ch-3-phase-diagram}. Although there is a relatively large distance between the three points, each exchange-correlation functional predicts that the LAO/STO interface is in region II, i.e. a single second order transition at a reduced critical thickness $\dphi$. The difference in LDA can be attributed to the shallower tilt double well. The difference between PBE and PBEsol appears to arise from the difference in the untilted dielectric constants.

We also performed calculations to investigate the effect of strain on the carrier transition, using the PBEsol exchange-correlation functional. Calculations were repeated but with small amounts of compressive and tensile strain applied to $a_{\text{STO}}$, up to $\pm$1\%. The results are included in Fig.~\ref{fig:ch-3-phase-diagram}, indicated by the smaller dots. We found that a compressive strain moved the point towards the origin and a tensile strain moved the point up and to the right, although it appears that a significantly larger amount of strain than the ones investigated here would be required to change the order of the carrier transition. It may be possible to change the character of the transitions using biaxial strain for a system with a point which lies closer to a boundary between regions, however.



%

\end{document}